\documentclass[aps,twocolumn]{revtex4}

\pdfoutput=1

\usepackage{eurosym}
\usepackage{amsfonts}
\usepackage{amsmath}
\usepackage{amssymb,epsf}
\usepackage{color}
\usepackage{epstopdf}
\usepackage{graphicx}
\usepackage{float}
\usepackage{caption}
\usepackage{subfig}
\usepackage{hyperref}
\usepackage{mathtools}
\usepackage{lipsum}

\definecolor{aogreen}{rgb}{0.0, 0.5, 0.0}

\def\ketm#1{  \left\vert  #1   \right\rangle   }

\def\bram#1{  \left\langle  #1   \right\vert   }

\begin{document}
	
	\title{One-dimensional quantum walks driven by two-entangled-qubit coins}
	\author{S. Panahiyan$^{1,2,3}$ \footnote{%
			email address: shahram.panahiyan@uni-jena.de} and S. Fritzsche$^{1,2,3}$%
		\footnote{%
			email address: s.fritzsche@gsi.de} }
	
	\affiliation{$^1$Helmholtz-Institut Jena, Fr\"{o}belstieg 3, D-07743 Jena, Germany  \\
		$^2$GSI Helmholtzzentrum f\"{u}r Schwerionenforschung, D-64291 Darmstadt, Germany \\
		$^3$Theoretisch-Physikalisches Institut, Friedrich-Schiller-University Jena, D-07743 Jena, Germany}

\begin{abstract}
We study one-dimensional quantum walk with four internal degrees of freedom resulted from two entangled qubits. We will demonstrate that the entanglement between the qubits and its corresponding coin operator enable one to steer the walker's state from a classical to standard quantum-walk behavior, and a novel one. Additionally, we report on self-trapped behavior and perfect transfer with highest velocity for the walker. We also show that symmetry of probability density distribution, the most probable place to find the walker and evolution of the entropy are subject to initial entanglement between the qubits. In fact, we confirm that if the two qubits are separable (zero entanglement), entropy becomes minimum whereas its maximization happens if the two qubits are initially maximally entangled. We will make contrast between cases where the entangled qubits are affected by coin operator identically or else, and show considerably different deviation in walker's behavior and its properties.
\end{abstract}

\maketitle

\section{Introduction}

Quantum walks (QWs), i.e. walks driven by the laws of quantum mechanics, are known to behave very differently from their classical counterparts \cite{Kempe}. In contrast to classical walks (CWs), QWs exhibit a ballistic spread for its probability density distribution (PDD). Therefore, QWs have been found an efficient framework to develop new (quantum) algorithms \cite{Shenvi}, increase the processing power to solve computationally hard problems \cite{Ambainis3} and to simulate other quantum systems \cite{Mohseni}. QWs are also known as universal computational primitives \cite{Lovett} and generators of PDD \cite{Montero2017}. Therefore, they are ideal for quantum simulation \cite{Aspuru-Guzik}. In addition, these walks were used to explore topological phases \cite{Kitagawa,Panahiyan2019}, build neural networks \cite{Dernbach}, prepare quantum states \cite{Franco} and engineer them \cite{Innocenti}. Experimentally, QWs have been realized with ultracold atoms \cite{Karski}, photons \cite{Schreiber}, ions \cite{Zahringer}, Bose-Einstein condensate \cite{Dadras} and optical-network \cite{Barkhofen,Lorz}. In fact, one of the advantages of QW is the possibility of demonstrating it by means of different systems.

Entanglement is one of the resources of quantum systems which has no classical counterpart \cite{Horodecki}. This resource plays a crucial role in quantum information and its applications/protocols such as superdense coding \cite{Muralidharan}, teleportation \cite{Bouwmeester}, cryptography \cite{Ekert}, quantum computation \cite{Jozsa} and algorithmic construction \cite{Hirsch,Fillettaz}. For these reasons, there has been a growing interest to create entangled states and make them available for different applications \cite{Horodecki}. Naturally, it is possible to use the entanglement as a resource in QWs as well.

In this paper, we consider a single walker with four internal degrees of freedom (internal states) that moves within a one-dimensional position space. The four internal states are realized by two qubits that are initially entangled with each other and are driven by a tensor product of two single-qubit coin operators \cite{Venegas,Liu2009,Liu2012}. Previously, only special cases of such walk were investigated and it was found that: These walks exhibit a persistent major peak at the initial position and two other distinguishable peaks at the most right and left hand side positions \cite{Venegas}. Also, the probability of finding the walker at any given location eventually becomes stationary and non-vanishing \cite{Liu2009}. In this work, in contrast, we generalize coins and initial states. In particular, we show that: a) The walk could show rather diverse behaviors including: classical like, self-self trapped, perfect transfer with highest velocity, Two-, Three- and Four-peaks-zone. This makes entanglement between qubits a resource for introducing novel behaviors and obtain previously reported ones. b) We find zero probability density at specific locations for walker when single-qubit coin operators are not identical. c) We show that symmetry of PDD, the most probable place to find the walker and maximization (minimization) of its entropy depend on the initial entanglement between the qubits.

The structure of paper is as follows: First, we introduce the setup of the walk, its parameters and highlight some of its properties (\ref{Ing}). Next, we simulate the walk for two scenarios concerning the coin's structure with two different initial states. We investigate how coins' and initial states' parameters affect the walker's behavior and its properties (\ref{Effect}). Then, we study the evolution of entropy and extract the conditions for its maximization (minimization) (\ref{Ent}). The paper is concluded with some closing remarks in section \ref{Con}. In two appendix, we study the most generalized possible initial state for the QW \ref{Generalized} and provide analytical description for cases we have addressed in the paper \ref{Analytical}.

\section{Setup of the walk} \label{Ing}

The walker is a quantum system that has four internal states and moves step wise in one-dimensional position space. The internal states of the walker are given by two entangled qubits. Therefore, the Hilbert space of the coin (internal states), $\mathcal{H}_{C}$, is spanned by $\{ \ketm{00},\: \ketm{11},\: \ketm{10},\: \ketm{01} \}$. The coin operator of the walk is a tensor product of two single-qubit coin operators (sub-coins), $\widehat{C} =  \widehat{C}_{1} \otimes \widehat{C}_{2}$ where
\begin{eqnarray}
\widehat{C}_{1} & = & \cos \theta\: \ketm{0}_{C} \bram{0} \:+\:\sin \theta\: \ketm{0}_{C} \bram{1}                                                                           \notag
 \\[0.1cm]
&   & 
\quad +\:
\sin \theta\: \ketm{1}_{C} \bram{0} \:-\: \cos \theta\: \ketm{1}_{C} \bram{1} \:, \label{c1}
 \\[0.1cm]
\widehat{C}_{2} & = & \cos \gamma\: \ketm{0}_{C} \bram{0} \:+\:\sin \gamma\: \ketm{0}_{C} \bram{1}                                                                           \notag
\\[0.1cm]
&   & 
\quad +\:
\sin \gamma\: \ketm{1}_{C} \bram{0} \:-\: \cos \gamma\: \ketm{1}_{C} \bram{1} \: . \label{c2}
\end{eqnarray}

Both $\widehat{C}_{1}$ and $\widehat{C}_{2}$ can be understood as rotation matrices that are characterized by their rotation angles, $\theta$ and $\gamma$. We can decompose these sub-coins into Pauli-$X$ and -$Z$ gates as $\widehat{C}_{1}= \cos \theta \sigma_{Z} + \sin \theta \sigma_{X}$ and $\widehat{C}_{2}= \cos \gamma \sigma_{Z} + \sin \gamma \sigma_{X}$ where $\sigma_{Z}$ and $\sigma_{X}$ are Pauli matrices. The coin operator is then obtained as

\begin{eqnarray}
\widehat{C} & = &  
\ketm{00}_{C}(
\cos\theta\: \cos\gamma\: \bram{00}+\cos\theta\: \sin\gamma\: \bram{01}             \notag 
\\[0.1cm]                                                                           
 &   & 
\quad +\:                                                                           
\sin\theta\: \cos\gamma\: \bram{10}+\sin\theta\: \sin\gamma\: \bram{11}             
) \quad +\:                                                                         \notag
\\[0.1cm]                                                                           
&   & 
\ketm{01}_{C}(
\cos\theta\: \sin\gamma\: \bram{00}-\cos\theta\: \cos\gamma\: \bram{01}             \notag
\\[0.1cm]
&   & 
\quad +\:
\sin\theta\: \sin\gamma\: \bram{10}-\cos\theta\: \sin\gamma\: \bram{11}             \notag
) \quad +\:                                                                          
\\[0.1cm]                                                                           
&   & 
\ketm{10}_{C}(
\sin\theta\: \cos\gamma\: \bram{00}+\sin\theta\: \sin\gamma\: \bram{01}             \notag
\\[0.1cm]        
&   & 
\quad -\:
\cos\theta\: \cos\gamma\: \bram{10}-\cos\theta\: \sin\gamma\: \bram{11}             \notag
) \quad +\:                                                                         \notag 
\\[0.1cm]                                                                           \notag  
&   &                                                                               \notag
\ketm{11}_{C}(
\sin\theta\: \sin\gamma\: \bram{00}-\sin\theta\: \cos\gamma\: \bram{01} \\[0.1cm]
&   & 
\quad -\:
\cos\theta\: \sin\gamma\: \bram{10}+\cos\theta\: \cos\gamma\: \bram{11}),                                                                                 \label{coin}                          
\end{eqnarray}
where we call $\theta$ and $\gamma$ coin's parameters. The considered sub-coins are unitary and the coin operator is consequently unitary. If $\widehat{C}$ acts upon the internal states of the walker, it results into a superposition of internal states. It should be noted that in our setup, there is only one walker that performs the quantum walk and this is different from the case where two entangled particles, hence two entangled walkers, are involved in the process of the quantum walk \cite{Singh}.

The walker moves along a one-dimensional position space where its Hilbert space, $\mathcal{H}_{P}$, is spanned by $\{ \ketm{x}_{P}: x\in \mathbb{Z}\}$. The conditional shift operator that moves the walker is given by 
	
\begin{eqnarray}
\widehat{S} & = & 
\ketm{00}_{C} \bram{00} \otimes \sum \ketm{x+1}_{P} \bram{x}   \quad +\:            \notag 
\\[0.1cm]
&   & 
(\ketm{10}_{C} \bram{10}  + 
 \ketm{01}_{C} \bram{01}) \otimes \sum \ketm{x}_{P} \bram{x}     \quad +\:            \notag 
\\[0.1cm]
&   & 
\ketm{11}_{C} \bram{11} \otimes \sum \ketm{x-1}_{P} \bram{x},                        \label{shift}
\end{eqnarray}
in which due to internal states, at each step, the walker can move to right, left or remain in the same position (similar to shift operators in Ref. \cite{Inui2004,Inui2005}). This is sketched schematically in Fig. \ref{Fig0}.

\begin{center}
	\centering
	\includegraphics[width=0.85\linewidth]{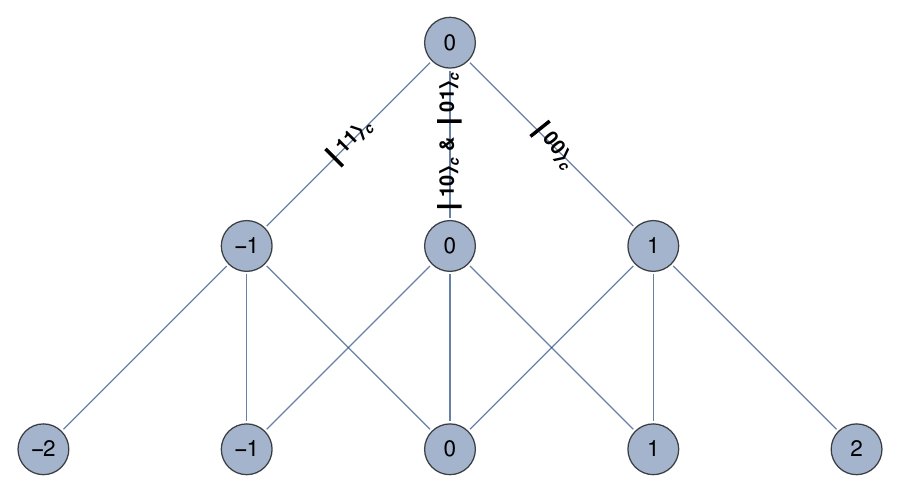}
	\captionof{figure}{Schematic plot for considered shift operator \eqref{shift} in the first step. The shift operator conditionally moves the walker to right, left or keeps it at the same position.} \label{Fig0}
\end{center}

The Hilbert space of the QW is given by $\mathcal{H} \:\equiv\: \mathcal{H}_{P} \otimes \mathcal{H}_{C}$ and the QW is the result of $T$ times successive application of the shift-coin operator on an initial state of the walker

\begin{eqnarray}
\ketm{\psi_{j}}_{Fin} & = &  \widehat{U}^{\:T}\: \ketm{\psi_{j}}_{Int} =  [\widehat{S} \widehat{C}]^{\:T}\: \ketm{\psi_{j}}_{Int}.
\end{eqnarray} 

Inspired by Bell states, in this paper, we consider two classes of initial states given by

\begin{eqnarray}
\ketm{\psi_{1}}_{Int} & = & (\cos\eta  \ketm{00}_{C} \:+\: e^{I\phi}\sin\eta \ketm{11}_{C})\otimes \ketm{0}_{P},                                                                     \label{int1}
\\[0.3cm]
\ketm{\psi_{2}}_{Int} & = & (\cos\alpha \ketm{10}_{C} \:+\: e^{I\phi}\sin\alpha \ketm{01}_{C})\otimes \ketm{0}_{P}.                                                                     \label{int2}
\end{eqnarray}

The parameters $\eta$ and $\alpha$ of the initial states specify the amount of initial entanglement between the two qubits \cite{Venegas}. $\phi$ is a phase factor that controls how different internal states of initial state should interfere with each other through the walk and the complete isolation of the interfere achieved by $\phi= \pi/2$. Note that for $\phi= 0 \text{ and } \pi$ with $\eta=\alpha=\pi/4$ in Eqs. \eqref{int1} and \eqref{int2}, we obtain the Bell states that are maximally entangled. Schematically, Fig. \ref{Fig00} shows how superpositions are created for three different initial states of \eqref{int1}, \eqref{int2} and \eqref{int3} if the coin operator \eqref{coin} is applied on them.

\begin{center}
	\centering
	\includegraphics[width=0.95\linewidth]{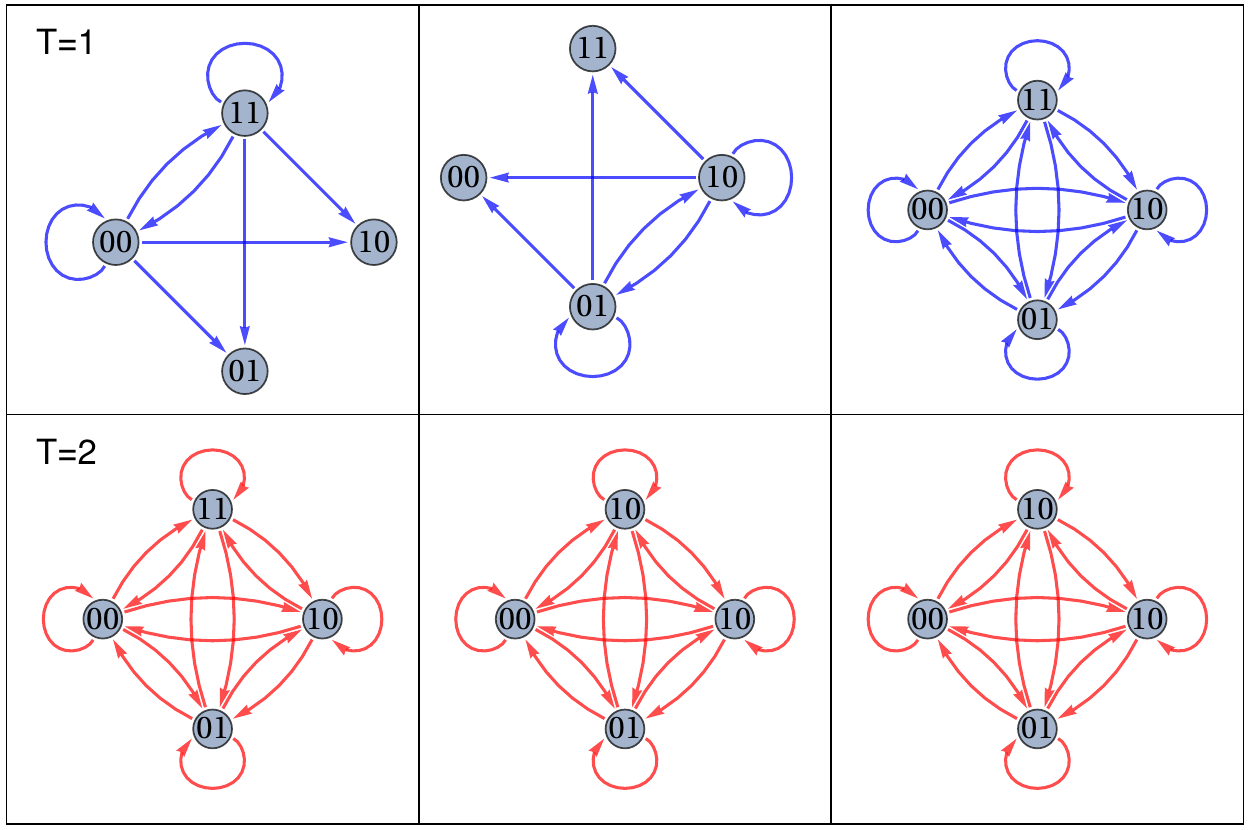}
	\captionof{figure}{Schematic plot for considered coin operator \eqref{coin} in the first two subsequent steps for three initial states of \eqref{int1} (left column), \eqref{int2} (middle column) and \eqref{int3} (right column). The application of the coin operator creates a superposition of all internal states.} \label{Fig00}
\end{center}

Our quantum walk has four degrees of freedom, Therefore, there are four conditional shifting for the walker based on four degrees of freedom. Out of these four conditional shifting, two of them keeps the walker at the same position in each step. At first glance, it seems that due to considered coin \eqref{coin} and shift operators \eqref{shift}, the probability density of initial position would be the highest through the evolution of the QW. This was reported in Refs. \cite{Venegas,Liu2009}. In contrast, we will show that this is not always the case and the walk exhibits significantly different behavior depending on the coin's and initial state's parameters.

Before we proceed, we define a few terms to make the upcoming explanations more clear. Two-peaks-zone corresponds to the case where PDD of the walker in position space has two major peaks in it. Accordingly, Three- and Four-peaks-zone indicate the existence of three and four major peaks. A Gaussian PDD is called classical like behavior. Complete localization takes place when probability density in a specific position is unit while zero for other positions. The extreme zones correspond to most left and right hand sides positions. 

The coin operator is made out of two sub-coins. If the rotation angles for these two sub-coins are identical, then both qubits are modified identically through the walk. In contrast, if they are not identical, then the two qubits are modified at different ratio. We call these two cases \textit{identical sub-coins} and \textit{non-identical sub-coins}, respectively. The Pauli-$X$ and $-Z$ gates are where the sub-coins are Pauli matrices of $\sigma_{X}$ and $\sigma_{Z}$, respectively.

\section{Effects of coin's and initial state's parameters on walk} \label{Effect}

Next, we investigate how the coin and the initial state affect the evolution of the QW. To this end, we build the wave function of the walker by iterations of the unitary operator of the coin-shift on the initial states of the walker in the most general form. We consider two cases of identical sub-coins ($\theta = \gamma$) and non-identical ones ($\theta \neq \gamma$). For both cases, the walk is analyzed for the initial states given in Eqs. \eqref{int1} and \eqref{int2}. In section \ref{Discussion}, we discuss the physical interpretation of our results. We limit our study to $\eta, \theta, \phi, \alpha, \beta \in [0,\pi/2]$. Tables \ref{T1} and \ref{T2} give summary regarding possible PDD that can be observed for identical and non-identical sub-coins. 

\subsection{Identical sub-coins: $\theta = \gamma$} \label{identical}

Generally, we observe Three- and Four-peaks-zone in PDD for both of the initial states except for four specific cases (see Fig. \ref{Fig1}). For sufficiently large $\theta$ ($\theta \rightarrow \pi/2$), the sub-coins approximate to Pauli-$X$ gates and the observed PDD would become Gaussian. When the sub-coins becomes Pauli-$X$ gates ($\theta = \pi/2$), the PDD completely localizes in starting position or two positions around it. This complete localization remains intact through the evolution of the walk which indicates that the walker is self-trapped \cite{Buarque}. In contrast, for sufficiently small $\theta$ ($\theta \rightarrow 0$), the sub-coins approximate to Pauli-$Z$ gates and the type of PDD differs between the two initial states (Two-peaks-zone for \eqref{int1} and Three-peaks-zone for \eqref{int2}). In case of Pauli-$Z$ gates for sub-coins ($\theta = 0$), we have significantly different behaviors for the two initial states. For initial state of \eqref{int1}, the PDD localizes equally in two mobile positions and in the mobility of the localizations, we have perfect transfer with highest velocity to both left and right side positions for the walker. Whereas, for initial state of \eqref{int1}, the walker becomes completely self-trapped at the starting position or two positions around it. Therefore, we observe that the internal states used for initial state alongside of rotation angles (sub-coins) determine the type of the PPD and walker's behaviors. This is in contrast to the simple QW with two internal states in which the type of PDD is independent of the internal states in initial state of the walker.

The amplitudes of internal states in initial state (initial entanglement between the qubits) affect PDD differently. For initial state of Eq. \eqref{int1}, the probability densities of the right (left) hand positions are decreasing (increasing) functions of $\eta$ while central probability density remains approximately independent of it (see Fig. \ref{Fig1}). In addition, the symmetrical PDD is obtained when initial state is maximally entangled ($\eta = \pi/4$). In contrast, in initial state of Eq. \eqref{int2}, the symmetry of PDD is independent of $\alpha$, hence initial entanglement between the qubits, and only the values of probability densities are affected by it. In fact, the central probability densities are decreasing functions of initial entanglement between the qubits ($\alpha$) while the extreme zones probabilities are increasing functions of it. Finally, we observe that the effectivity of variation in phase parameter ($\phi$) depends on internal states used in initial state. While for initial state of Eq. \eqref{int1}, $\phi$ insignificantly affects the probability densities in each position, in initial state of Eq. \eqref{int2}, variation of $\phi$ significantly modifies central probability densities.

The variance of the walk is determined by the sub-coins and step number, while it is independent of internal states in the initial state of the walker (see Fig. \ref{Fig1}). For Pauli-$X$ and $Z$ gates, the variance is almost zero. Otherwise, the variance is a decreasing function of the rotation angle, $\theta$. On the other hand, as walk proceeds, the variance monotonically increases. It should be noted that the type of the PDD is independent of amplitudes of internal states in the initial state (hence $\eta$ and $\alpha$), the evolution of the QW and the phase factor (hence $\phi$).

Before we finish this section, we provides details why classical like behavior (Gaussian PDD) is observable for our quantum walks. The probability density of a Gaussian distribution is given by
\begin{equation}
f(x\mid \mu, \sigma^2)= \frac{e^{-\frac{(x-\mu)^2}{2\sigma^2}}}{\sqrt{2 \pi \sigma^2}},
\end{equation}
in which $\mu$ is the mean or expectation of the distribution, $\sigma$ is the standard deviation, and $\sigma^2$ is the variance. In Fig. \ref{Fig121}, it is evident that we can completely match some of the probability density distributions of our quantum walks with Gaussian distributions. Therefore, we can draw the conclusion that in such cases, the probability density distributions of the quantum walks are classical like.

\begin{table*}
  \caption{\label{T1} Type of probability density distribution as a function of coin's rotation angles; $\theta = \gamma$. The initial states' parameters change the amplitudes of the probability density in positions but does not change the generic picture of type of distribution.}
  \begin{ruledtabular}
  \begin{tabular}{cccc}
	$\theta$               & $\ketm{\psi_{1}}_{Int}$ & $\ketm{\psi_{2}}_{Int}$  \\ \hline
	$\theta = 0$           & \textit{Localized in two positions} & \textit{Localized in one position}  \\
	$\theta \rightarrow 0$ & \textit{Two-peaks-zone} & \textit{Three-peaks-zone}  \\
	$\theta \rightarrow \pi/2$ & \textit{Gaussian} & \textit{Gaussian} \\
	$\theta = \pi/2$           & \textit{Localized in one position} & \textit{Localized in one position} \\
			\textit{Otherwise}         & \textit{Three- and Four-peaks-zone} & \textit{Three- and Four-peaks-zone} \\
  \end{tabular}
  \end{ruledtabular}
\end{table*}

\begin{table*}
  \caption{\label{T2} Type of probability density distribution as a function of coin's rotation angles; $\theta \neq \gamma$. The initial states' parameters change the amplitudes of the probability density in positions but does not change the generic picture of type of distribution.}
  \begin{ruledtabular}
  \begin{tabular}{cccc}
	$\theta$               & $\ketm{\psi_{1}}_{Int}$ & $\ketm{\psi_{2}}_{Int}$  \\ \hline
	$\theta \rightarrow \pi/2$ & \textit{Two-peaks-zone} & \textit{Two-peaks-zone} \\
	$\theta = \pi/2$           & \textit{Two-peaks-zone} & \textit{Two-peaks-zone} \\
	$\theta \rightarrow \gamma$           & \textit{Three-peaks-zone} & \textit{Three-peaks-zone} \\
	$\theta = \gamma$           & \textit{Three-peaks-zone} & \textit{Three-peaks-zone} \\			
	\textit{Otherwise}         & \textit{Four-peaks-zone} & \textit{Four-peaks-zone} \\
  \end{tabular}
  \end{ruledtabular}
\end{table*}

\subsection{Non-identical sub-coins: $\theta \neq \gamma$} \label{Nonidentical}

In this section, we consider different rotation angles ($\theta \neq \gamma$) for the sub-coins. This indicates that entangled qubits building up the coin space are modified differently in each step.

\begin{figure*}[!htbp]
	\centering
	\subfloat[]
	{\begin{tabular}[b]{c}%
		\includegraphics[width=0.52\linewidth]{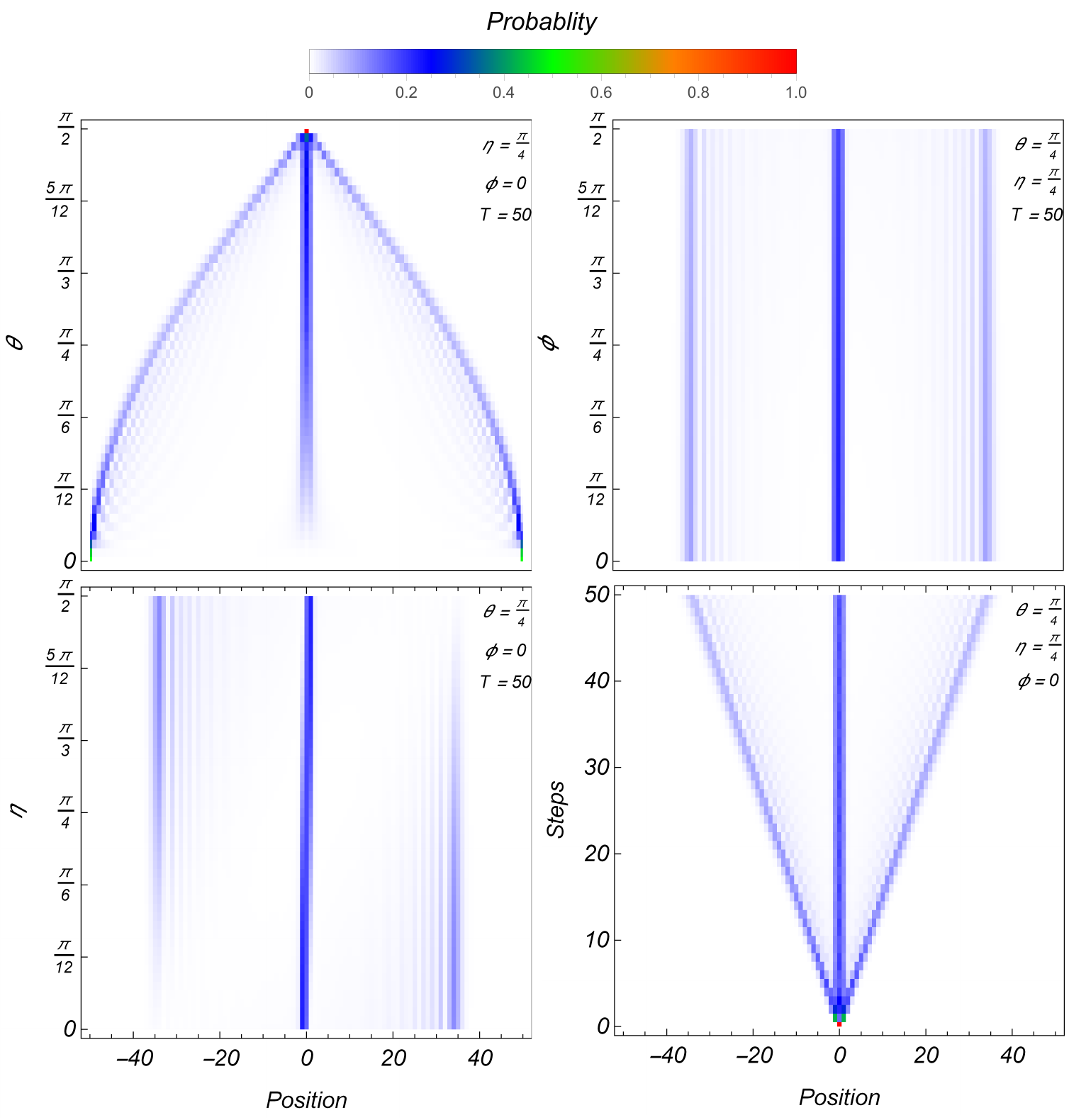} 	   		
	\end{tabular}}
	\subfloat[]
	{\begin{tabular}[b]{c}%
		\includegraphics[width=0.52\linewidth]{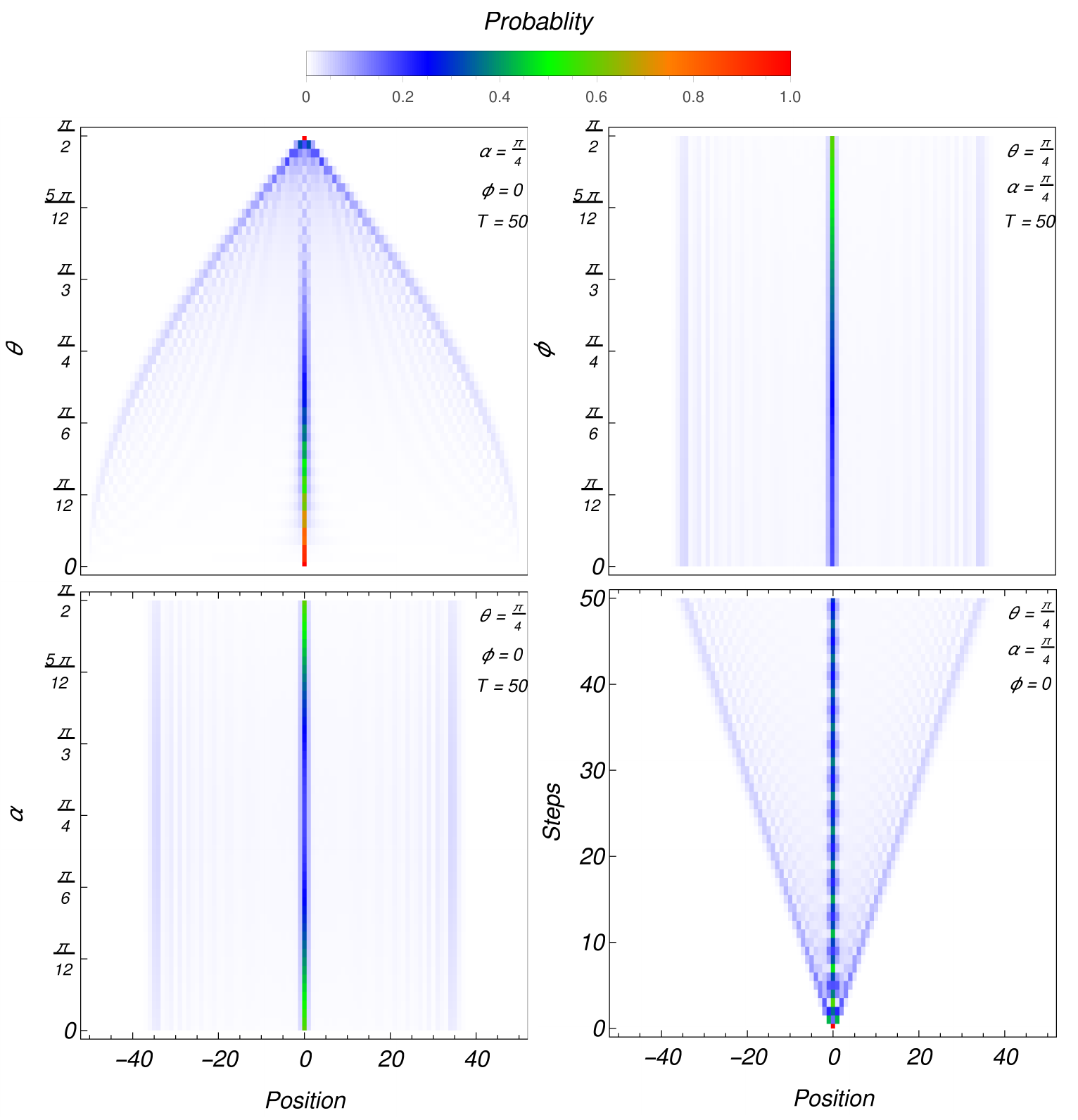}			
	\end{tabular}}	
	\caption{Identical sub-coins ($\theta = \gamma$): Probability density distribution for initial states $\ketm{\psi_{1}}_{Int}$ (a) and $\ketm{\psi_{2}}_{Int}$ (b). In (a), we observe that symmetrical probability density distribution happens when the inital state is maximally entangled. In addition, if sub-coins approximate to Pauli-$X$ gates we have Gaussian distribution and for Pauli-$X$ gates, the walker becomes completely self-trapped. In contrast, for Pauli-$Z$ gate sub-coins, the walker has two mobile probability densities which have perfect transfer with highest velocity to both left and right side positions for the walker. In (b), the entanglement in initial state has no effect on symmetry of PDD and it only changes the place where it is more probable to find the walker. As for sub-coins, the only significant differences is when sub-coins become Pauli-$Z$ gate in which walker becomes self-trapped (contrary to perfect transfer for the other initial state).} \label{Fig1}
\end{figure*}

\begin{figure*}[!htbp]
	\centering
	\subfloat[]	
	{\begin{tabular}[b]{cc}%
		\includegraphics[width=0.4\linewidth]{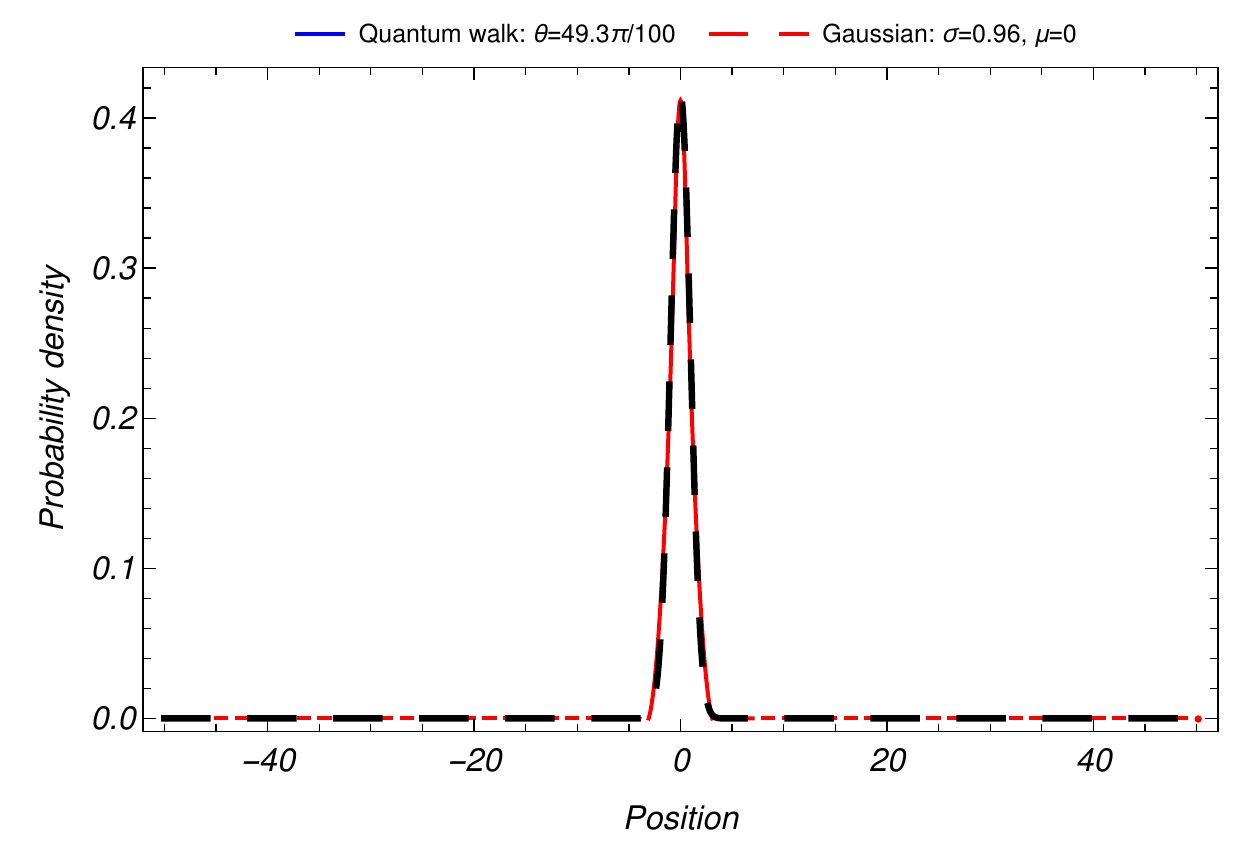}							   	
		\end{tabular}}
		\subfloat[]
		{\begin{tabular}[b]{cc}%
		\includegraphics[width=0.4\linewidth]{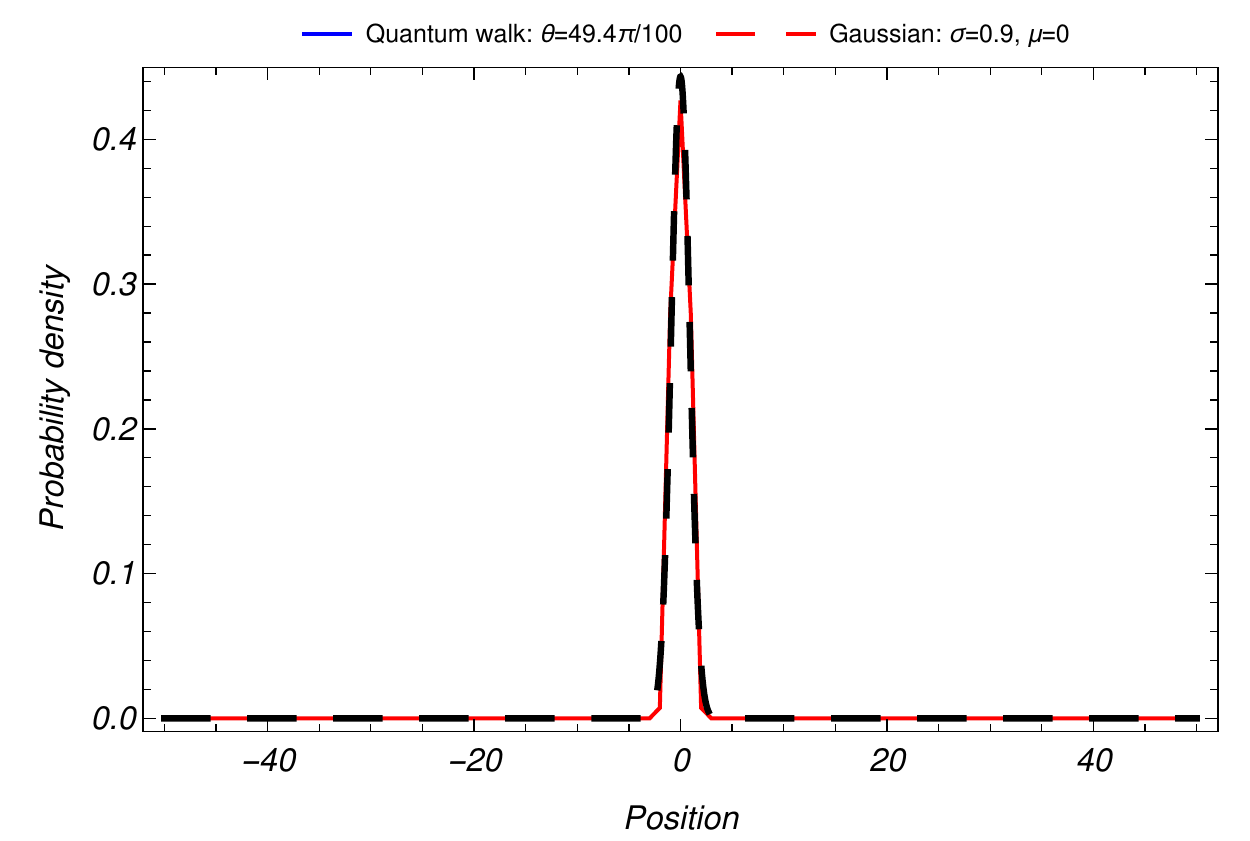}						   	
	\end{tabular}}		
	\caption{Probability density distribution with $\alpha=\eta=\pi/4$ and $\phi=0$ for $50th$ of the quantum walks (solid lines) for two initial states of $\ketm{\psi_{1}}_{Int}$ (a) and $\ketm{\psi_{2}}_{Int}$ (b). The dashed lines are Gaussian (normal) distributions with $x=0$ and different parameters. We observe that the probability density distributions resulted from quantum walks can be matched by Gaussian distributions. This enables us to draw conclusion that quantum walks have classical like behaviors in these cases.} \label{Fig121}
\end{figure*}

\begin{figure*}[!htbp]
	\centering
	\subfloat[]
	{\begin{tabular}[b]{c}%
		\includegraphics[width=0.52\linewidth]{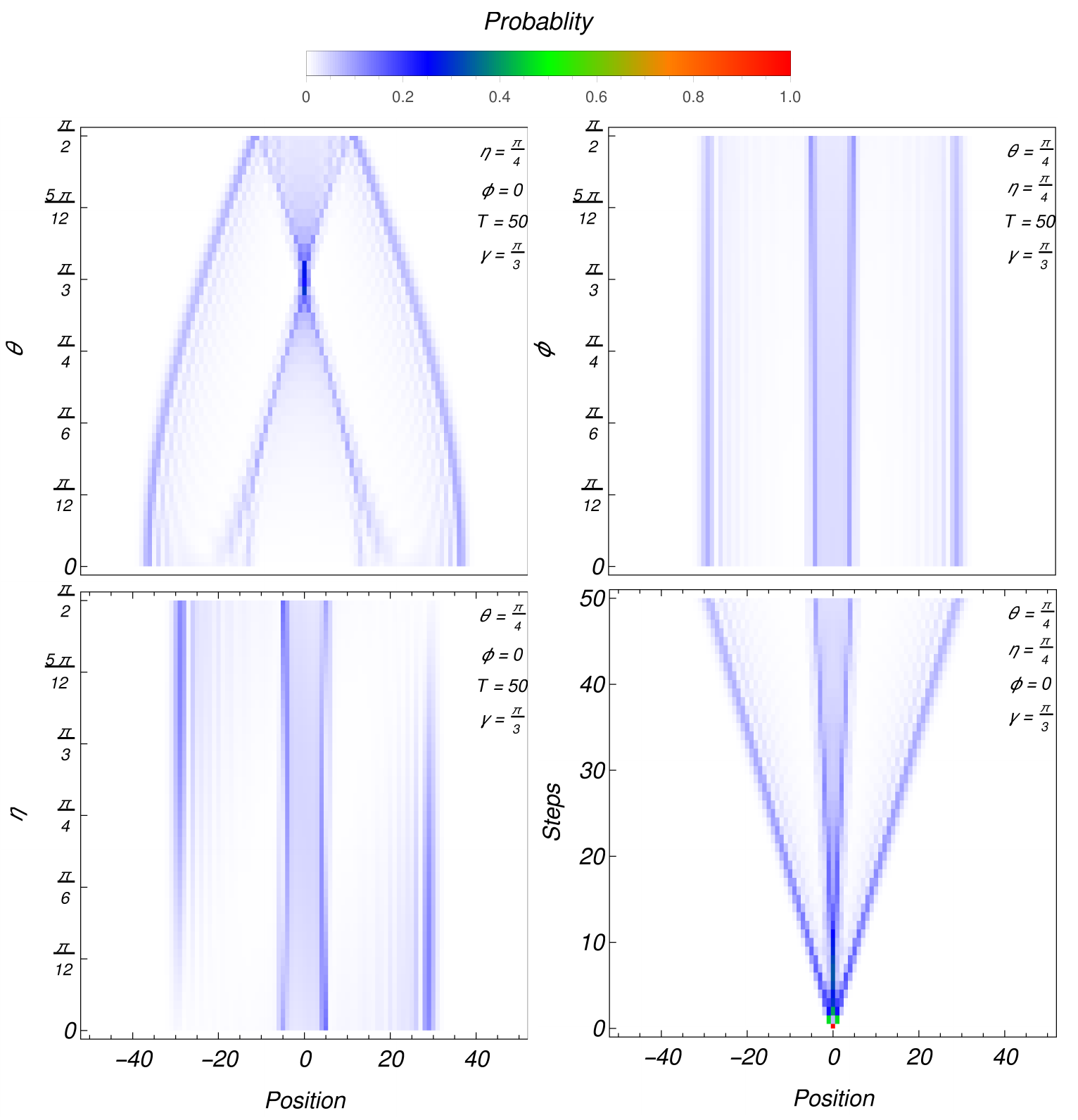} 		
	\end{tabular}}
	\subfloat[]
	{\begin{tabular}[b]{c}%
		\includegraphics[width=0.52\linewidth]{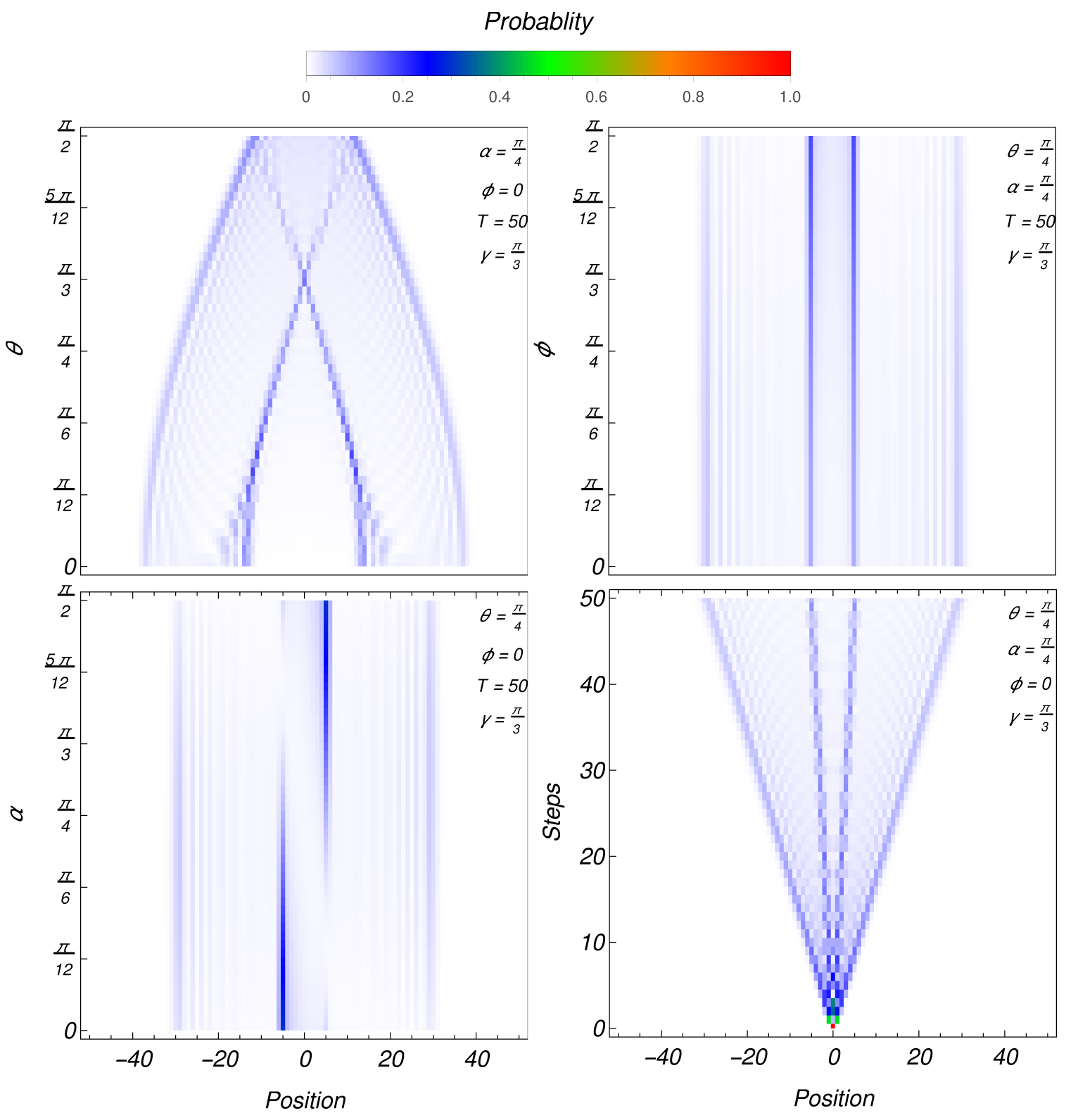}			
	\end{tabular}}	
	\caption{Non-identical sub-coins ($\theta \neq \gamma$): Probability density distribution for initial states $\ketm{\psi_{1}}_{Int}$ (a) and $\ketm{\psi_{2}}_{Int}$ (b). For both (a) and (b), we see that self-trapped, perfect transfer and Gaussian distribution are omitted. The only types of probability density distributions that are observed Two- , Three- and Four-peaks-zone. The symmetry for probability density distribution for both initial depends on initial entanglement between the qubits and happens if initial states are maximally entangled ($\eta=\alpha=\pi/4$). Finally, if one of the sub-coins approximates to or become a Pauli-$Z$ gate, a formation of a region with almost zero probability density between two inner peaks takes place.} \label{Fig2}
\end{figure*}

The significant differences between identical and non-identical sub-coins are the absences of localizations and Gaussian distributions in PDD for non-identical sub-coins (compare Figs. \ref{Fig2} with \ref{Fig1}). For approximately Pauli-$X$ and Pauli-$X$ gate, we observe the Two-peaks-zone in PDD for both of the initial states. In contrast, if one of the sub-coins approximates to Pauli-$Z$, Pauli-$Z$ gate, $\theta > \gamma$ and $\theta < \gamma$, Four-peaks-zone are formed in PDD. In limiting case of $\theta \rightarrow \gamma$ and $\theta = \gamma$, the PDD is Three-peaks-zone type. Therefore, the QW is no longer able to provide Classical like behavior (Gaussian distribution) in its PDD and self-trapped is vanished as a possible characteristic behavior for the walker. Additionally, contrary to previous case, we find that the type of PDD becomes independent of the internal states in initial state of the walker.

Modification from identical sub-coins to non-identical sub-coins significantly changes the effects of amplitudes of internal states in initial state (initial entanglement between the qubits) on PDD (see Fig. \ref{Fig2}). The symmetrical PDD is only conditioned to initial entanglement between the qubits (independent of phase factor and rotation angles) and we observe it if the qubits are maximally entangled ($\eta=\alpha=\pi/4$). While the left (right) hand side probability densities are almost decreasing (increasing) function of $\eta$ (exception is some of central peaks), the opposite is observed for variation of $\alpha$. On the other hand, The variation of phase factor ($\phi$) moderately affects central probability densities in PDD for both of initial states while it has rather weaker effects on other probability densities in PDD.

The variance of the walk is dominated only by the sub-coins and step number (see Fig. \ref{Fig2}). Similar to QW with identical sub-coins, the variance is a decreasing function of the rotation angle, $\theta$ and it has monotonical increment through the QW. The type of PDD could change as walk proceeds for some special cases of sub-coins where $\theta \rightarrow \gamma$ but generally, the type is independent of the evolution of the walk. Finally, if one of the sub-coins approximates to or becomes a Pauli-$Z$ gate, a region between the inner peaks with almost zero probability densities is formed. This region of zero-probability-density is an increasing function of the steps.


\subsection{Discussion on results} \label{Discussion}

Identical sub-coins, hence identical modification of entangled qubits through the QW, results into diverse PDD for the walker's wave function where the classical like behavior (Gaussian distribution) is the least expected one. Previously, it was pointed out that classical like behavior in QWs could be obtained by introduction of decoherence into the walk \cite{Brun,kosik,Kendon,Romanelli,Annabestani,Venegas-Andraca,Alberti} or using a step-dependent coin \cite{Panahiyan}. Here, we see that using two entangled coins is an additional method to simulate classical like behavior. In the case of decoherence, the classical behavior is obtained when entanglement between coin and position spaces is omitted. Whereas here, the entanglement between two qubits is used as a tool to obtain such behavior.

The classical like behavior is a feature raised by the presence of entangled qubits and not just a feature of standard coined QW \cite{Strauch}. In our study, the classical like behavior was observed when sub-conis are identical. As soon as the sub-conis become non-indetical, the classical like behavior is completely omitted. This shows that the classical like behavior is rooted in whether entangled qubits are modified identically or differently through the QW. In addition, in the entropy diagrams (see Figs. \ref{Fig5} and \ref{Fig6}), we observe that the classical behavior are found for the entropy/entanglement which is very small and close to the minimum. Therefore, the classical like behavior happens when entropy becomes very small and falls into specific range. These two points show that classical like behavior observed here is different from the one showed in Ref. \cite{Strauch} and is due to the entangled qubits.

The Two-peaks-zone is similar to usual one-dimensional walk with two internal degrees of freedom. The Three-peaks-zone is also reported for QWs with decoherence \cite{Alberti} or with step-dependent coin \cite{Panahiyan}. The Four-peaks-zone is uniquely observed for the setup considered in this paper and is one of the characterization of walk with two entangled qubits. It should be noted that Arnault et al, in Ref. \cite{Arnault} introduced a family of QWs on the line with discrete $U(N)$ gauge invariance. In their work, they obtained Four-peaks-zone PDD. But their result is different from Four-peaks-zone obtained here. First of all, $4$-dimensional coin given in their paper is not due to two entangled qubits. Their coin space is built by one qubit with four internal degrees of freedom. In addition, due to shift operator of the mentioned paper, in odd(even) steps, only odd(even) positions have non-zero probability densities. Whereas in our work, in each step, both odd and even positions have non-zero probability densities. Therefore, the Four-peaks-zone of the present work and the mentioned paper are different. 

As the sub-coins become different ($\theta \neq \gamma$), the QW's behavior and its properties change significantly. The first noticeable issue is the absence of classical like behavior and complete localization in walker's PDD (compare tables \ref{T1} and \ref{T2}). The second issue is modification in effectivenesses of internal states (specially $\ketm{10}_{C}$ and $\ketm{01}_{C}$) on symmetry of PDD (initial state being maximally entangled or else) and probability densities in each position. Finally, the variance of the PDD is another factor that highly depends on whether sub-coins are identical or non-identical. The mixing and hitting times are determined by the coin operator of the QW, hence sub-coins. Therefore, the identicality or else of the sub-coins plays major role in the applications such as development of algorithms \cite{Childs2004,Santha}.

The special cases of QW introduced here, could be employed to observe a genuine Parrondo’s paradox \cite{Rajendran,Machida}. In Parrondo’s paradox, it is stated that two losing games can produce a winning outcome. This paradox has been adapted in many physical and biological systems to explain how these systems work. In Ref. \cite{Rajendran}, it was also pointed out that such QW could demonstrate a quantum ratchet. Ref. \cite{Machida} provides some limit laws for the special cases of QW studied here. It was proved that by the means of these limit laws (these analytical tools), one can study some of the Parrondo type behaviors and explore the “phase space” of parameters of a winning quantum game. Another important application of our QW is in state engineering and/or preparations for entangled systems. It is well known that preparing desired entangled system is rather a difficult task \cite{Kovlakov}. Recently, it was shown that QWs could be used for state engineering and preparation purposes \cite{Innocenti,Giordani}. Our setup here provides large family of PDD. In addition, it is possible to use the parameters provided in our setup to determine desirable entropy/entanglement for both walker and its entangled qubits (see sections \ref{Effect} and \ref{Ent}). Therefore, it is possible to use our setup to prepare entangled qubits with specific entanglement and characterizations.

\section{Evolution of Entropy} \label{Ent}

Here, we investigate the modification of entropy present in the state of walker. The goal is to understand the effects of different parameters on walker's entropy. This is done by investigating the entropy as a function of steps, coin's and initial state's parameters.

\subsection{von Neumann entropy} 

In CW and information theory, the entropy of a discrete PDD is investigated by Shannon entropy \cite{Shannon,Nielsen}. As for quantum physics, the von Neumann approach is usually used. This is because for open quantum systems (non-pure states), the density matrix formalism is employed to study system's evolution. Accordingly, the entropy should also be calculated by the properties of the density matrix. The density matrix at step $T$ of the walk is given by 

\begin{eqnarray}
\hat{\rho}_{T} & =  \ketm{\psi}_{T} \prescript{}{T}{\bram{\psi}}.  &  
\end{eqnarray}

\begin{figure*}[!htbp]
	\centering
	\subfloat[]	
	{\begin{tabular}[b]{cc}%
		\includegraphics[width=0.52\linewidth]{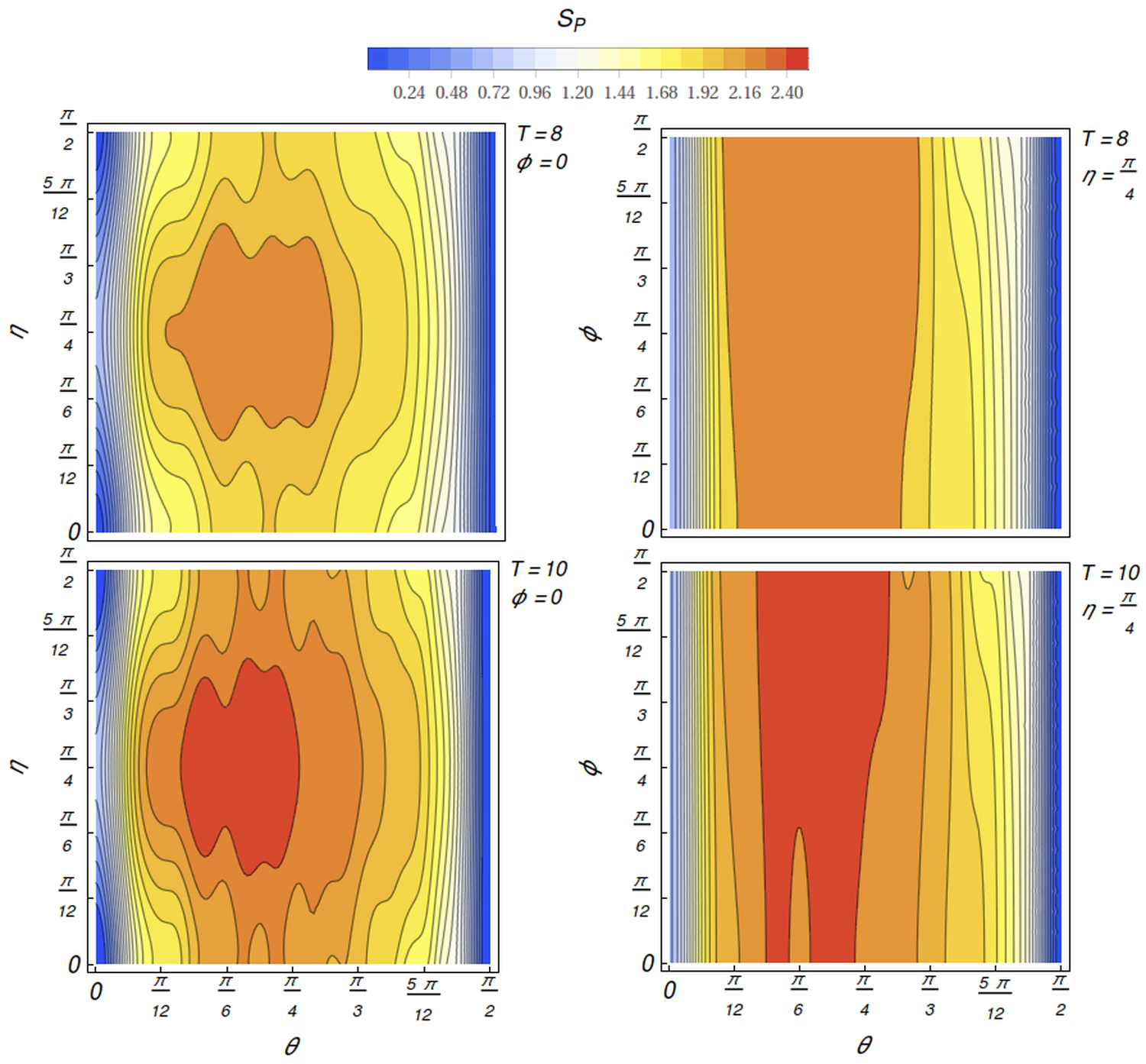}							   	
	\end{tabular}}
	\subfloat[]	
	{\begin{tabular}[b]{cc}%
		\includegraphics[width=0.52\linewidth]{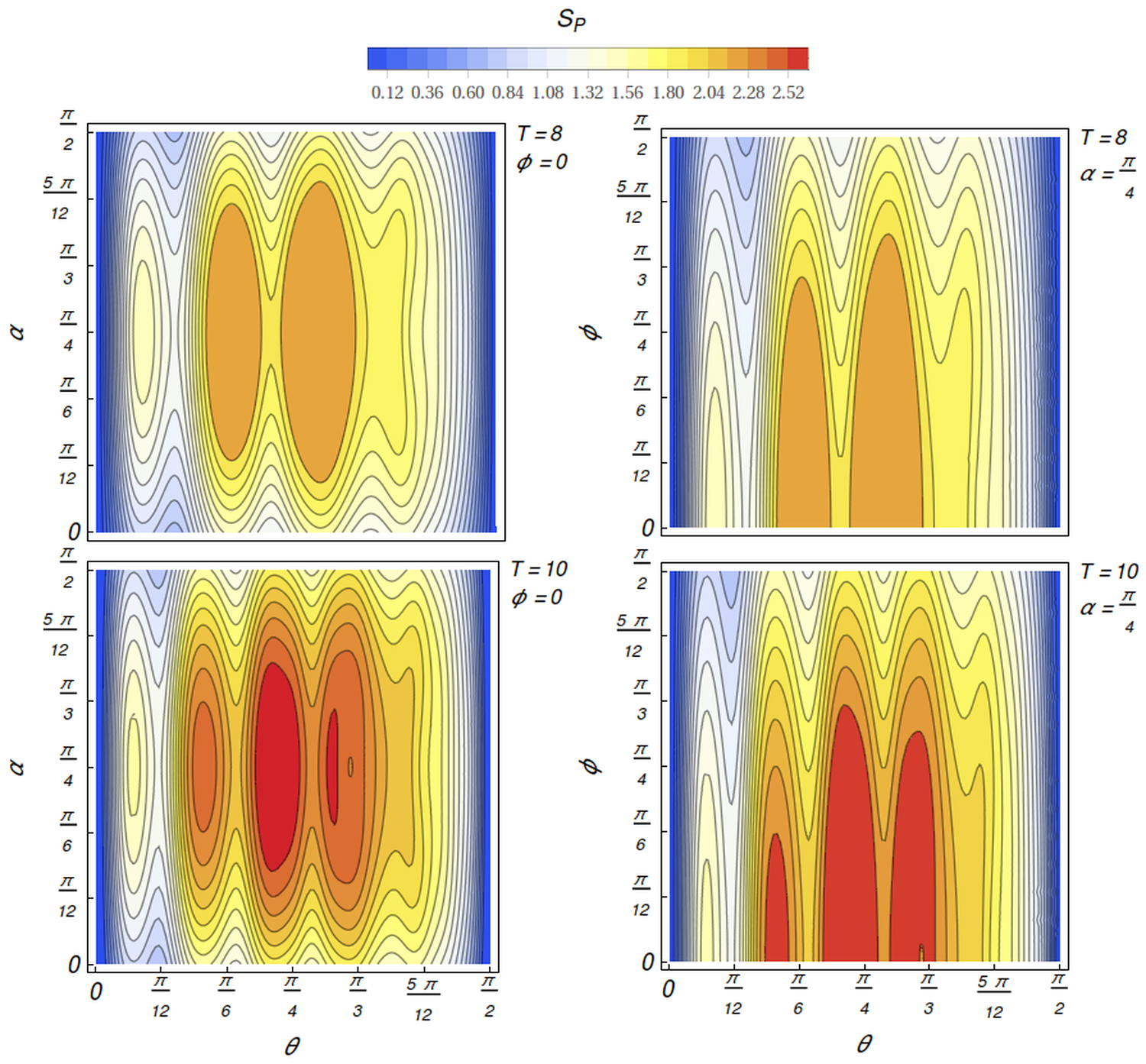}		
		
	\end{tabular}}		
	\caption{Identical sub-coins ($\theta = \gamma$): Entropy of position space as a function of coin's and initial state's parameters for steps $8$ and $10$ with initial states $\ketm{\psi_{1}}_{Int}$ (a) and $\ketm{\psi_{2}}_{Int}$ (b). In (a) and (b), we see that minimum of entropy is found when sub-coins approximate to Pauli-$X$ and $Z$ gates or become these gates, and the entanglement between the two qubits becomes zero (two qubits become separable). In contrast, the entropy maximizes when the initial entanglement between the qubits becomes maximum, hence inital states are maximally entangled. The entropy's behavior is symmetrical with respect to $\eta=\alpha=\pi/4$. Variation of phase factor has small effect on entropy for initial state of $\ketm{\psi_{1}}_{Int}$ while moderate effect for $\ketm{\psi_{2}}_{Int}$.} \label{Fig5}
	\end{figure*}

\begin{figure*}[!htbp]
	\centering
	\subfloat[]	
	{\begin{tabular}[b]{cc}%
		\includegraphics[width=0.52\linewidth]{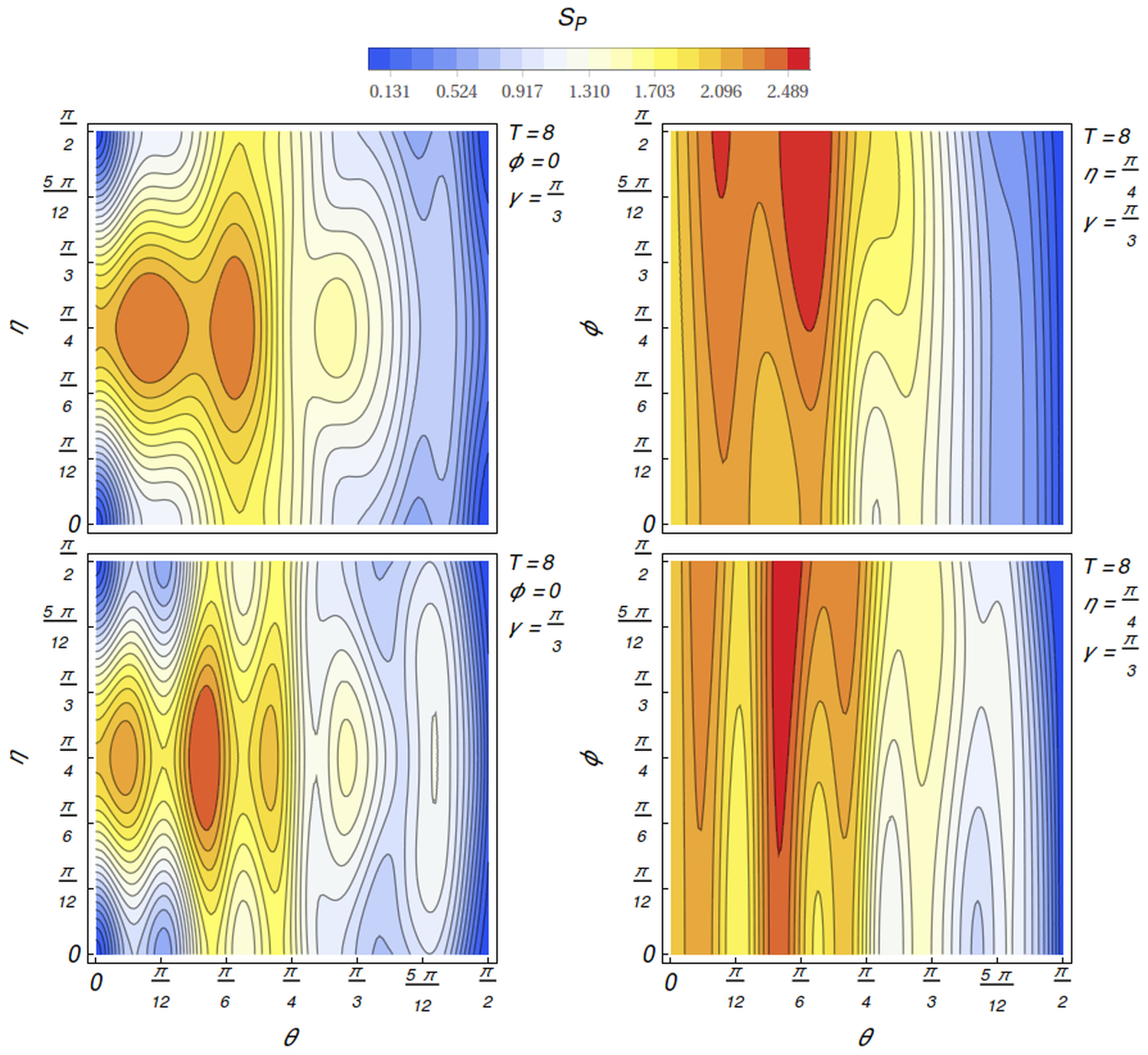}							   	
	\end{tabular}}
	\subfloat[]	
	{\begin{tabular}[b]{cc}%
		\includegraphics[width=0.52\linewidth]{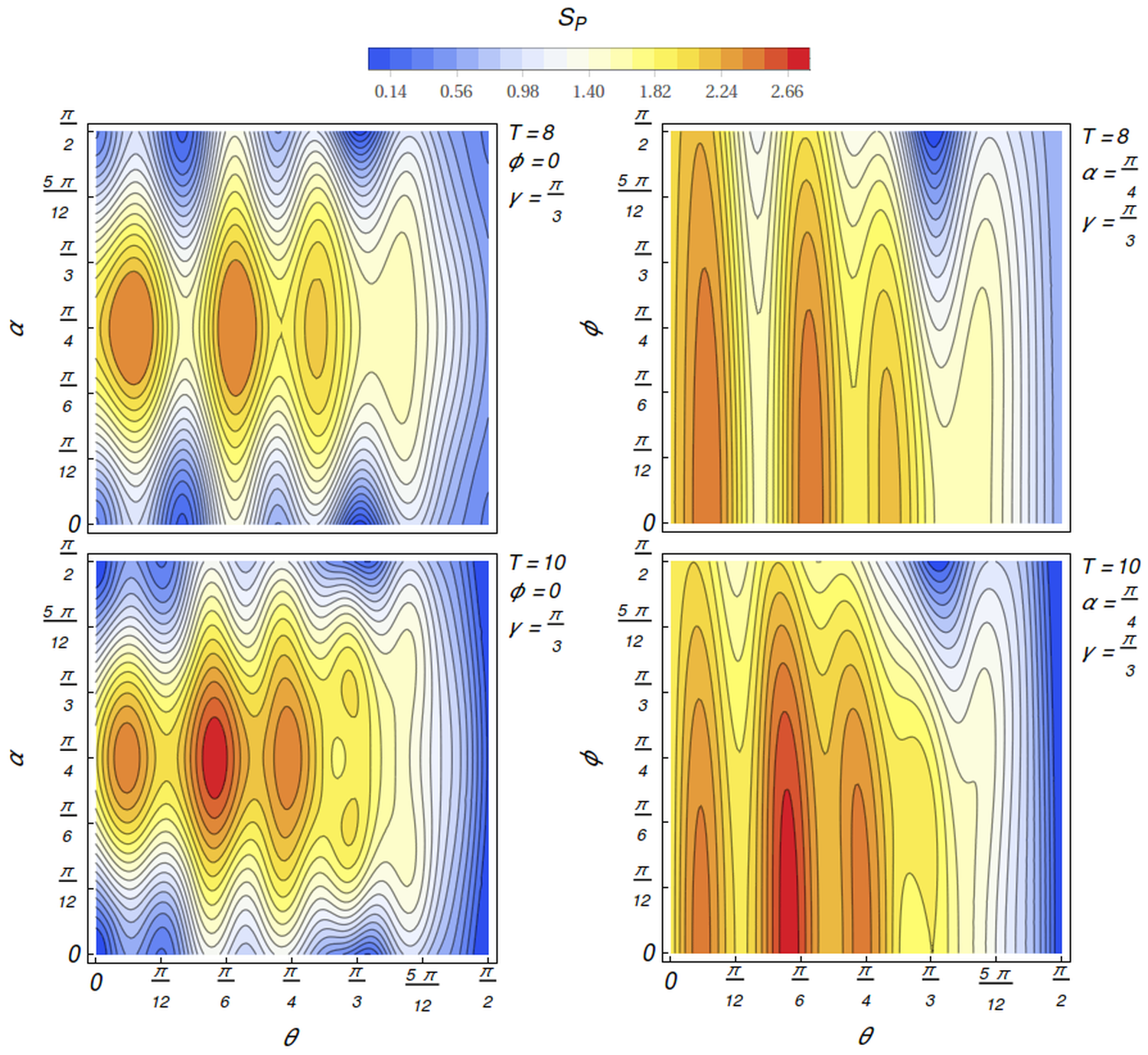}						   	
	\end{tabular}}		
	\caption{Identical sub-coins ($\theta \neq \gamma$): Entropy of position space as a function of coin's and initial state's parameters for steps $8$ and $10$ with initial states $\ketm{\psi_{1}}_{Int}$ (a) and $\ketm{\psi_{2}}_{Int}$ (b). In (a) and (b), we see that entropy minimizes for one of the sub-coins approximates to Pauli-$X$ and $Z$ gates or becomes these gates, and the two qubits become separable (zero entanglement). In contrast, the entropy maximizes if the qubits are maximally entnagled in the initial state of the walker. The entropy's behavior is symmetrical with respect to $\eta=\alpha=\pi/4$. Variation of phase factor has considerably stronger effect comparing to identical sub-coins for both initial states of $\ketm{\psi_{1}}_{Int}$ and $\ketm{\psi_{2}}_{Int}$.} \label{Fig6}
\end{figure*}

The von Neumann method uses the reduced density matrix, $\hat{\rho}^{P}_{T}=\text{Tr}_{C}(\hat{\rho}_{T})$, to calculate the entropy of position space \cite{Ide}. The von Neumann entropy at time $T$ is given by 

\begin{eqnarray}
S_{P} &=& -\text{Tr}(\hat{\rho}^{P}_{T} Log\hat{\rho}^{P}_{T}),  \label{von}
\end{eqnarray}
where for walk under consideration in this paper, it yields

\begin{eqnarray}
S_{P} &=& -\sum_{n}P_{n,T}LogP_{n,T},  \label{entrop1}
\end{eqnarray}
in which, $P_{n,T}$ are eigenvalues of Hermitian matrix with the element $\hat{\rho}^{P}_{T}$. For pure states, the Shannon and von Neumann entropies become identical. Therefore, $P_{n,T}$ is the probability density of the position $n$ at step $T$. The results for initial states of \eqref{int1} and \eqref{int2}, with two cases of identical and non-identical sub-coins are plotted in Figs. \ref{Fig5} and \ref{Fig6} (for position space. We limit our investigation to $\eta, \phi, \alpha, \theta \in [0, \pi/2]$.


\subsection{Identical versus Non-identical sub-coins} 

Overall, the entropy of position space increases as walk proceeds irrespective of choices for initial state and sub-coins (see Figs. \ref{Fig5} and \ref{Fig6}). The only exceptions are where localization takes place in which the entropy is zero (physically expected). The variation of entropy as a function of walker's parameters shows different properties for two cases of identical and non-identical sub-coins. 

For identical sub-coins, irrespective of the choice for initial state, the entropy minimizes when sub-coins approximate to Pauli-$X$ and $Z$ gates or become these gates and initial entanglement between the qubits is zero (see Fig. \ref{Fig5}). In contrast, the maximization of entropy happens if initial state of the QW is maximally entangled ($\eta=\alpha=\pi/4$) and the rotation angles of the sub-coins approach $\pi/4$, indicating that the sub-coins approximate to Hadamard gates. The entropy's behavior is symmetric with respect to $\eta=\alpha=\pi/4$. The entropy is not a sensitive function of phase factor, $\phi$, for initial state of \ref{int1} while its variation moderately modifies entropy for initial state of \ref{int2}. 

In case of non-identical sub-coins (see Fig. \ref{Fig6}), similarly, the entropy becomes minimum if at least one of the sub-coins approaches to Pauli-$X$ and $Z$ gates or become these gates and initial entanglement between the qubits vanishes (see Figs. \ref{Fig5}). The maximization of entropy is conditioned on initial state being maximally entangled but the rotation angle of sub-coin should approximate to $\theta=\pi/6$. On the other hand, we observe that variation of the phase factor significantly modifies the entropy for both initial states. Finally, similar to identical sub-coins, entropy's behavior is symmetric with respect to maximization of initial entanglement between the qubits ($\eta=\alpha=\pi/4$).

The entropy highly depends on initial entanglement between the qubits that build the coin space. Though the entropy is a dynamical quantity as a function of steps, we find that if the qubits are separable, zero initial entanglement between the qubits, the entropy becomes minimum while for maximally entangled initial states, the entropy attains the highest value. The minimization of the entropy also conditioned to at least one of the sub-coins approximates to Pauli-$X$ and $Z$ gates or become these gates. Such choices for sub-coins result into whether localized PDD, self-trapped walker, classical like behavior in PDD or small variance (see Figs. \ref{Fig1} and \ref{Fig2}). Therefore, we see that minimization of entropy happens when PDD has a more deterministic behavior. 

The symmetries observed for initial state's parameters are indentical irrespective of sub-coins being identical or else. Persistence of symmetry associated to initial state's parameters shows that they are more efficient to be utilized in designing algorithms \cite{Shenvi,Lovett}, quantum simulations \cite{Mohseni} and setups with quantum information applications. This is because if in experimental setups for these purposes, the sub-coins incidentally become non-identical, the symmetrical properties of initial state's parameters remain valid. On the other hand, the modification in symmetrical properties could be used as a method to recognize problems in experimental setups. For applications such as exploring topological phases \cite{Kitagawa}, implementation of quantum circuits \cite{Qiang} and quantum state preparation/engineering \cite{Franco,Innocenti} the symmetrical properties of entropy and PDD play an important role. 

The maximization and minimization of the entropy is decided by two factors: I) homogeneity of the probability density distribution, II) variance of the walk. This means that the walk that has a more homogeneous probability density distribution on sufficiently large number of positions would attain maximum of entropy and vice versa for minimum entropy. Through our simulations, we confirmed that symmetrical probability density distribution is found for maximally entangled initial states. Additionally, we showed that for maximally entangled initial states, the probability density distribution would be more homogeneous (please see Figs. \ref{Fig1} and \ref{Fig2}, and also the analytical results). Therefore, irrespective of step number, the maximum entropy will be found for maximally entangled initial states and consequently the minimum of entropy would be observed for separable qubits. As for the coin's parameter effects, we showed that if sub-coins approximate or become Pauli-X and Z gates, the probability density distribution would be concentrated in one, two or several positions. Therefore, the minimization of entropy for each step happens for such rotation angles. In contrast, for the maximization, the sub-coins should be far enough from these two limiting cases which brings us to the vicinity of rotation angle being $\theta=\pi/4$. This is again valid for arbitrary steps.

\section{Conclusion} \label{Con}

In this paper, we investigated the one-dimensional QW with four internal degrees of freedom (coin space). The coin space was built up by entanglement of two qubits. The study was done for two distinguishable initial states where for specific values of their parameters, they would yield the Bell states. The coin operator of the walk was made by tensor product of two single-qubit coin. 

We confirmed that such system or more precisely entanglement between two qubits could be used as a resource for obtaining different PDD in position space including classical like behavior, self-trapped walker and novel one (Four-peaks-zone). In addition, we pointed it out that properties of walk such as symmetry of PDD, maximization (minimization) of entropy and amplitude of probability density in each position are functions of the amount of entanglement in the initial state of the walk. We also showed that the walker's behavior highly depends on the ratio at which entangled qubits are modified. If the ratio of their modifications through walk is different, the dependency of the walker's properties on initial state's and coin's parameters would change and some specific properties/behaviors are eliminated (for example the classical like behavior and self-trapped walker). These show that the entangled qubits and the amount of entanglement between them are possible means for state engineering, preparation and quantum simulation.

The importance of our investigation is to highlight the possibilities that are provided by entanglement in QW. So far, it was shown that entanglement is a unique resource belonging only to quantum systems and absent in classical ones. While the CW is highly celebrated for its applications in different branches of science, for the QW, we indeed have more resources at our disposal. These resources could be utilized for more applications specially the ones that their similar ones in CW do not exit. Here, we showed that by introduction of two entangled qubits, we easily cover the classical like behavior alongside of previously reported ones for QW and obtain even novel ones. Therefore, one can draw the conclusion that entanglement and multi-entangled qubits for a single walker provide us with more efficient frameworks comparing to a single walker with a single qubit.  

\section{Appendix A: Generalized initial state} \label{Generalized} 

In this section, we consider the initial state to be given by all possible internal states available to us

\begin{eqnarray}
\ketm{\psi_{3}}_{Int} & = & 
[e^{I\phi} \sin\beta (\cos\alpha  \ketm{10}_{C} \:+\: \sin\alpha \ketm{01}_{C}) \quad +\:  \notag 
\\[0.1cm]
&   & 
\cos\beta (\cos\eta  \ketm{00}_{C} \:+\: \sin\eta \ketm{11}_{C})]\otimes \ketm{0}_{P},                                                                      \label{int3}
\end{eqnarray}
in which, $\alpha$, $\eta$ and $\beta$ determine the amount of initial entanglement between the qubits and $\phi$ is a phase factor that indicate how different internal sates should interfere with each other through the QW. 

\begin{figure*}[!htbp]
	\centering
	\subfloat[]
	{\begin{tabular}[b]{c}%
		\includegraphics[width=0.51\linewidth]{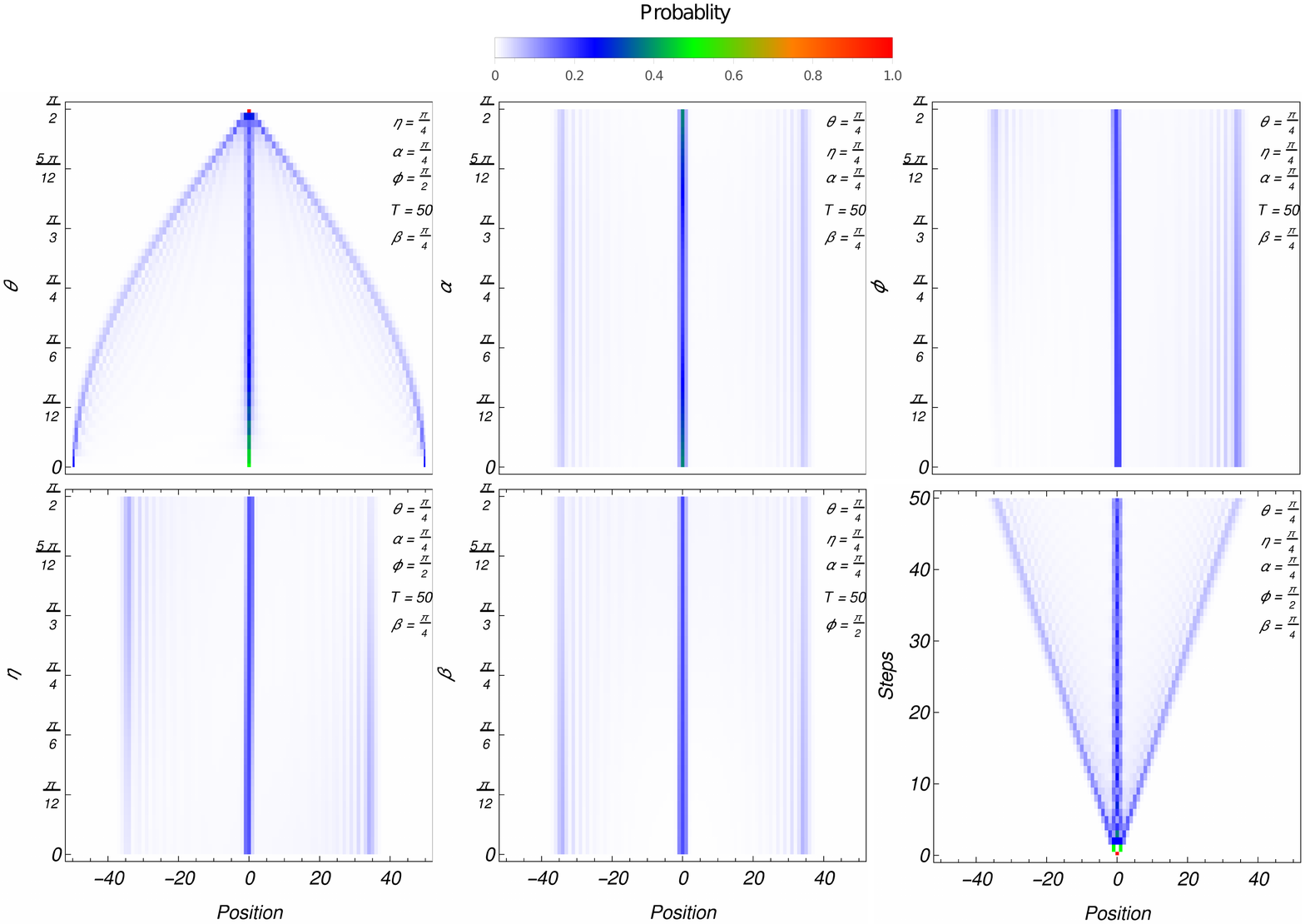} 		
	\end{tabular}}
	\subfloat[]
	{\begin{tabular}[b]{c}%
		\includegraphics[width=0.51\linewidth]{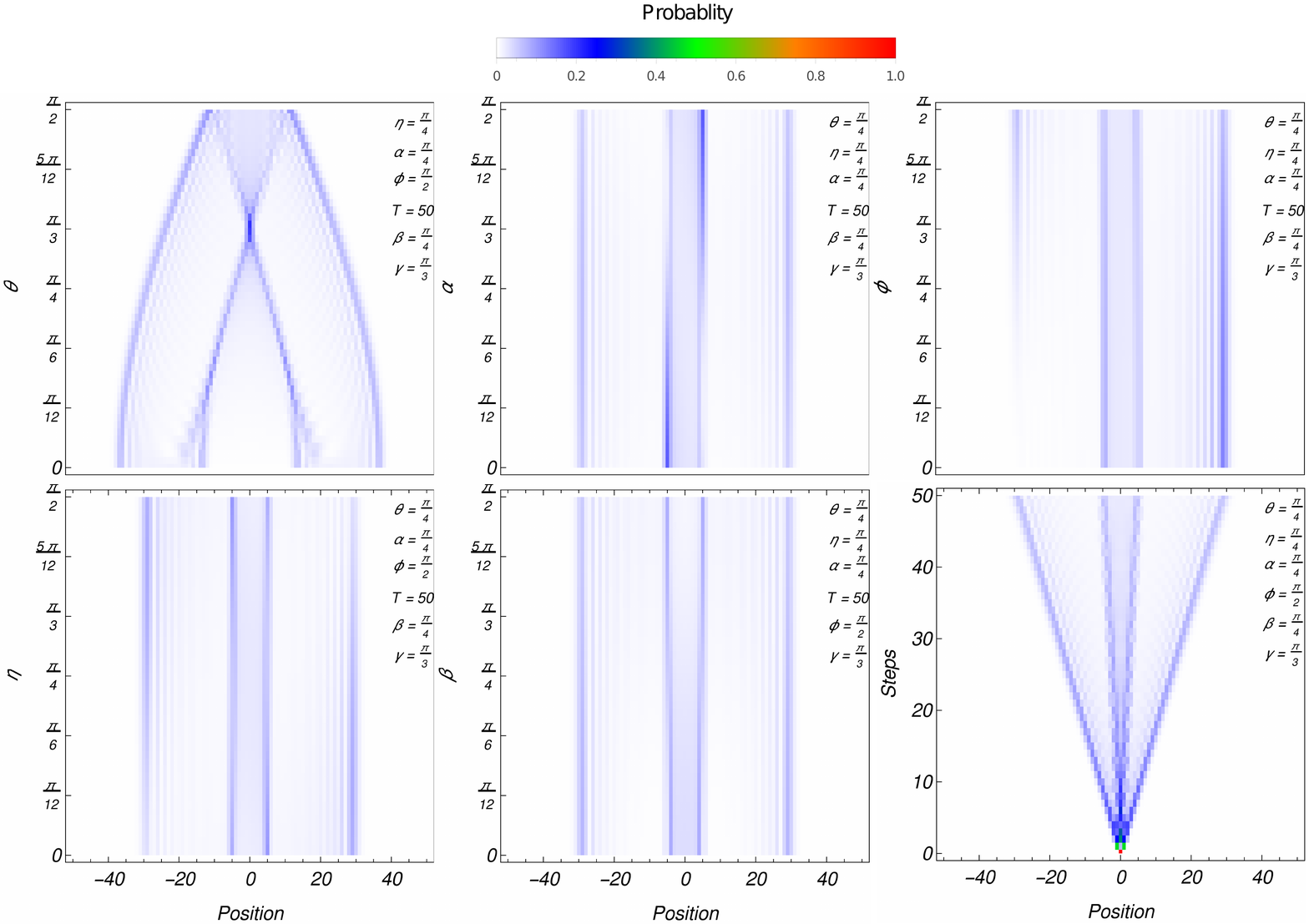}			
	\end{tabular}}	
	\caption{Probability density distribution for initial state $\ketm{\psi_{3}}_{Int}$ in case of identical sub-coins (a) and non-identical sub-coins (b). In (a), we observe that for sub-coins approximating to Pauli-$X$ gates, Gaussian distribution is formed in probability density distribution and self-trapped behavior for walker happens if the sub-coins are Pauli-$X$ gates. In contrast, for sub-coins that are Pauli-$Z$ gates, the walker partially becomes self-trapped while partially has perfect to right and left hand side positions. Modification to non-identical sub-coins (b) omits these behaviours for the walker. The other major differences between identical and non-identical sub-coins cases is the effectivity of certain internal states on symmetry of probability density distribution (compare the effects of variation of $\alpha$ in diagrams).} \label{Fig3}
\end{figure*}

For identical sub-coins (see Fig. \ref{Fig3}), if sub-coins approximate to Pauli-$X$ gates the classical like PDD is resulted and for Pauli-$X$ gate sub-coins, the walker would be self-trapped at the origin of position space or around it. In case of approximately Pauli-$Z$ gates for sub-coins, PDD starts to localized in three positions and such localization is completed when sub-coins are Pauli-$Z$ gates. Part of localization is at the origin while the other two positions of localization are mobile and have perfect transfer with highest velocity to both left and right side positions for the walker. Therefore, our walker is partially self-trapped while partially has perfect transfer. For other cases of sub-coins, Three-peaks-zone are formed in PDD (absence of Two- and Four-peaks-zone). 

On the other hand, we notice that the symmetrical PDD is obtained for $\phi=\pi/2$ and $\eta=\pi/4$ and it is independent of $\alpha$ and $\beta$. Therefore, the symmetry of PDD is independent of initial entanglements between the qubits and only depends on the amplitudes of specific internal states used in initial state. The variance of the walk is only determined by coin's parameters. It is worthwhile to mention that the left (right) hand side probability densities are increasing (decreasing) functions of $\eta$ and $\phi$. The central peak becomes minimum at $\alpha=\pi/4$ and it is an increasing function of $\beta$.

The modification from identical sub-coins to non-identical ones, makes $\alpha$ alongside of $\eta$ and $\phi$ a deterministic factor to have symmetrical PDD which happens when $\alpha=\eta=\pi/4$ and $\phi=\pi/2$ (see Fig. \ref{Fig3}). On the other hand, $\beta$ only changes the probability densities in each position. Therefore, being maximally entangled in initial state is not a necessary condition for having symmetrical PDD. Here, we also observe that if one of the sub-coins approximates to or becomes Pauli-$X$ gate, PDD would be Two-peaks-zone type. In contrast, for $\theta \leftarrow \gamma$ and $\theta = \gamma$, Three-peaks-zone are emerged in PDD. For other cases of sub-coin, we have Four-peaks-zone in PDD (see Table \ref{T3}).  

\begin{table*}
	\caption{\label{T3} Type of probability density distribution as a function of coin's rotation angles for initial state $\ketm{\psi_{3}}_{Int}$ .}
	\begin{ruledtabular}
		\begin{tabular}{cccc}
			$\theta$              & $\theta = \gamma$ & $\theta \neq \gamma$ \\ \hline
			$\theta \rightarrow \pi/2$ & \textit{Localized in three positions} & \textit{Two-peaks-zone} \\
			$\theta = \pi/2$           & \textit{Three-peaks-zone} & \textit{Two-peaks-zone} \\
			$\theta \rightarrow \gamma$           & \textit{Gaussian} & \textit{Three-peaks-zone} \\
			$\theta = \gamma$     & \textit{Localized in one position} & \textit{Three-peaks-zone} \\			
			\textit{Otherwise}    & \textit{Three-peaks-zone} & \textit{Four-peaks-zone} \\
		\end{tabular}
	\end{ruledtabular}
\end{table*}

\section{Appendix B: Generalized of sub-coins into $SU(2)$ ones}

\begin{figure*}[!htbp]
	\centering
	\subfloat[]
	{\begin{tabular}[b]{c}%
		\includegraphics[width=1\linewidth]{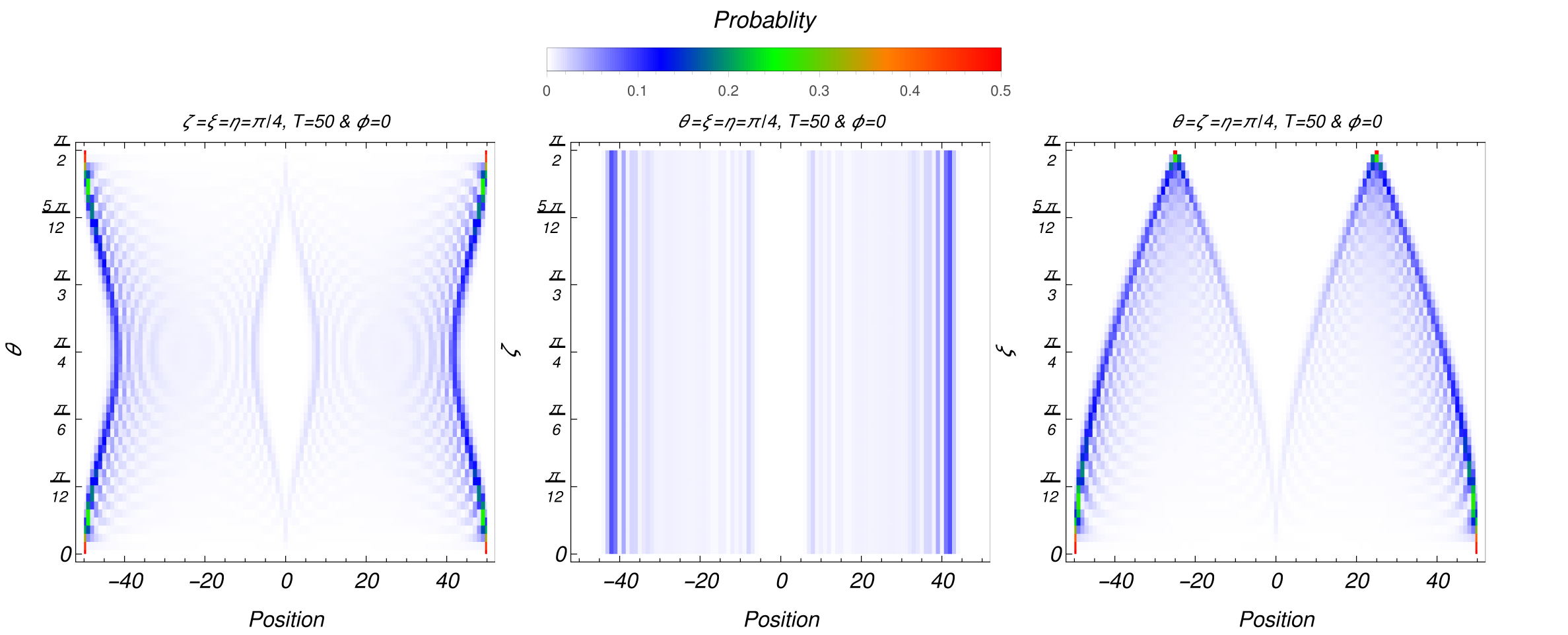} 		
	\end{tabular}}
	\\
	\subfloat[]
	{\begin{tabular}[b]{c}%
		\includegraphics[width=1\linewidth]{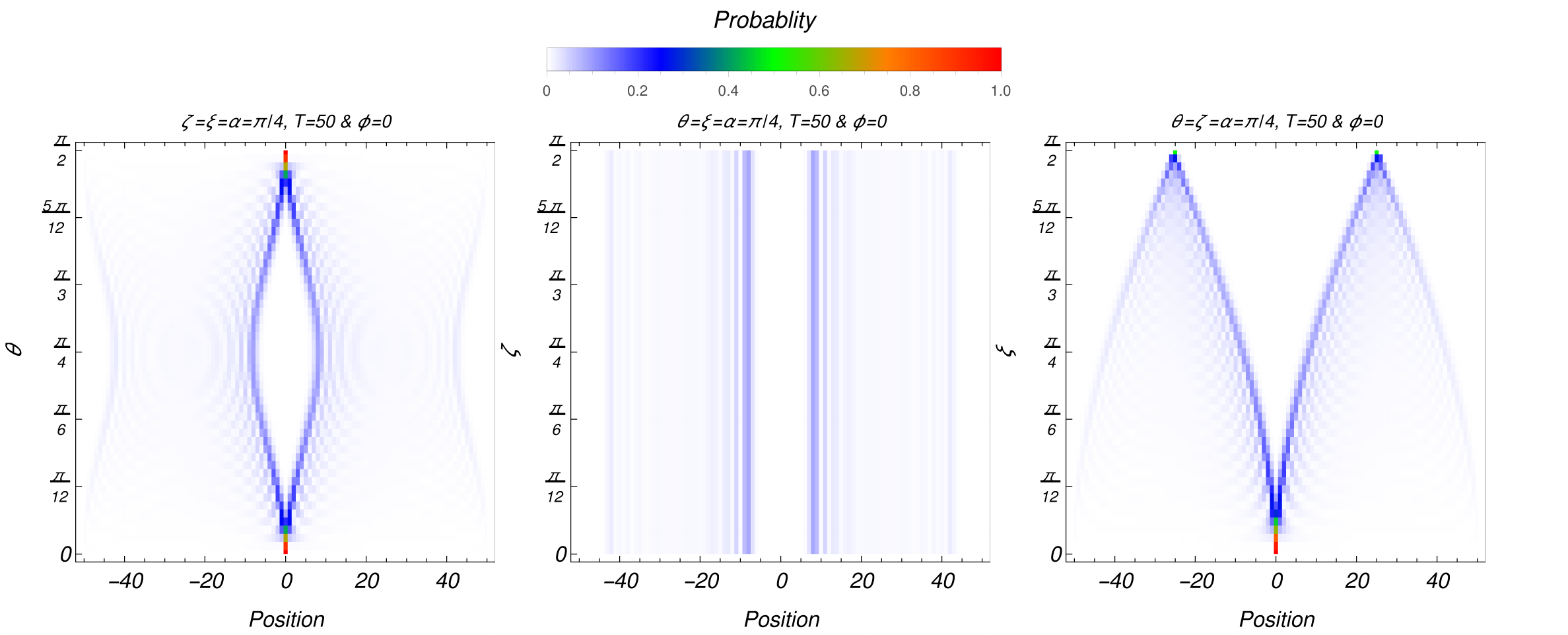} 		
	\end{tabular}}		
	\caption{Identical sub-coins ($\theta = \gamma$): Probability density distribution for initial states $\ketm{\psi_{1}}_{Int}$ (a) and $\ketm{\psi_{2}}_{Int}$ (b) with a generalized $SU(2)$ given in Eq. \eqref{c3}. Evidently, the generalization to $SU(2)$ results into elimination of the classical like and self-trapped behaviors for $\ketm{\psi_{1}}_{Int}$ whereas they remain present for initial state of $\ketm{\psi_{2}}_{Int}$. The coin parameters $\theta$ and $\xi$ give us further control over the probabilities in each position. Whereas the coin parameter $\zeta$ apparently has very insignificant effect on the properties of the walk.} \label{Fig22}
\end{figure*}

Next, we should address the possibility of generalizing sub-coins into $SU(2)$ in the setup of the walk. In Ref. \cite{Chandrashekar2008}, Chandrashekar et al. considered a generalized structure for the coin operator as 

\begin{eqnarray}
\widehat{C}_{1} & = & e^{I \xi} \cos \theta\: \ketm{0}_{C} \bram{0} \:+\: e^{I \zeta} \sin \theta\: \ketm{0}_{C} \bram{1}                                                                           \notag
\\[0.1cm]
&   & 
\quad +\:
e^{-I \zeta} \sin \theta\: \ketm{1}_{C} \bram{0} \:-\: e^{-I \xi} \cos \theta\: \ketm{1}_{C} \bram{1} \:, \label{c3}
\end{eqnarray}
and showed that by tuning the coin's parameter, one is able to maximize the variance of the quantum walk and improve the mixing time. Later, a scheme for Parrondo’s game was developed using the quantum walk with such a coin and it was shown that two individual players that lose the game can come up with a strategy to emerge as joint winners \cite{Chandrashekar2011}. In here, We question whether employing such coin would affect the symmetry of the PDD and or the probability densities in each position. We limit our study to identical coins with initial states of \eqref{int1} and \eqref{int2}. The results are presented in diagrams in Fig. \ref{Fig22}.

Evidently, the symmetry of the PDD is independent of the coin's parameters for both of the initial states. In contrast, the type of the PDD, variance and probability density in each position highly depends on $\theta$ and $\xi$, while apparently independent of $\zeta$. In the case of non-zero values for all three parameters of the coin, the classical like and self-trapped behaviors would be eliminated for initial states of \eqref{int1} whereas they remain present for initial sate of \eqref{int2}. For both of the initial states, the additional coin parameters give higher level of control over the probability densities in each position as well as the the type of the PDD and its variance. The only exception is $\zeta$ which for both cases seems has rather a very insignificant effect on walker's behavior.

\section{Appendix B: Analytical description} \label{Analytical}

In this section, we give an analytical description for our QWs studied here. In general, QW protocol indicates that the evolution of walk could be given by  
\begin{eqnarray}
\ketm{\psi_{j}}_{T} = U \ketm{\psi_{j}}_{T-1}= U^{2}\ketm{\psi_{j}}_{T-2}=...=U^T \ketm{\psi_{j}}_{0}.
\end{eqnarray}

Based on this, one can write the the amplitude of wave function of the walk at step $T$ as $\Psi_{T}(x)\equiv[\phi_{T}(x,1),\phi_{T}(x,2),\phi_{T}(x,3),\phi_{T}(x,4)]^{\zeta}$ where four components at position $x$ and step $T$ correspond respectively to the coin states $\ketm{00}, \: \ketm{11}, \: \ketm{10} \text{and} \: \ketm{01}$ and $\zeta$ denotes the transpose operator. Since our walker starts at the origin of position, \eqref{int1}-\eqref{int3}, the initial state will be obtained as $\Psi_{0}(0)=[\phi_{0}(0,1),\phi_{0}(0,2),\phi_{0}(0,3),\phi_{0}(0,4)]^{\zeta} \equiv [\beta_{1},\beta_{2},\beta_{3},\beta_{4}]^{\zeta}$ where $\sum^{4}_{z=1} |\beta_{z}|=1$.

The spatial discrete Fourier transform of $\Psi_{T}(x)$ is given by  

\begin{eqnarray}
\Psi_{T}(k) = \sum_{x\in \mathbb{Z}}\Psi_{T}(x)e^{ikx},
\end{eqnarray}
where for amplitudes of initial state, one can obtain 
 
\begin{eqnarray}
\Psi_{0}(k) = \Psi_{0}(0).
\end{eqnarray}

The evolution of walk will be then 
\begin{eqnarray}
\Psi_{T}(k) =U(k)^{T} \Psi_{0}(k),
\end{eqnarray}
in which 
\begin{eqnarray*}
U(k) & = & U_{1}(k/2) \otimes  U_{2}(k/2)=
\\
&   & 
\begin{bmatrix}e^{ik/2} & 0 \\0 & e^{-ik/2}\end{bmatrix} \widehat{C}_{1} 
\otimes \begin{bmatrix}e^{ik/2} & 0 \\0 & e^{-ik/2}\end{bmatrix} \widehat{C}_{2}.
\end{eqnarray*}

It is straight forward to find eigenvalues of $ U_{1}(k/2)$ and $U_{2}(k/2)$, respectively, as 

\begin{eqnarray*}
\lambda_{1,\pm} & = & \pm \sqrt{1-\cos^{2} \theta \sin^{2} (\frac{k}{2})}+i \cos \theta \sin (\frac{k}{2}),
\\
\lambda_{2,\pm} & = & \pm \sqrt{1-\cos^{2} \gamma \sin^{2} (\frac{k}{2})}+i \cos \gamma \sin (\frac{k}{2}).
\end{eqnarray*}

Since $U(k)=U_{1}(k/2) \otimes  U_{2}(k/2)$, the eigenvalues of $U(k)$ are obtained as 

\begin{eqnarray}
\left\{
\begin{array}{c} 
\Lambda_{1}= \lambda_{1,+} \lambda_{2,+}
\\[0.4cm]
\Lambda_{2}= \lambda_{1,+} \lambda_{2,-}
\\[0.4cm]
\Lambda_{3}= \lambda_{1,-} \lambda_{2,+}
\\[0.4cm]
\Lambda_{4}= \lambda_{1,-} \lambda_{2,-}
\end{array}  
\right.  ,                  
\end{eqnarray}
and the eigenvectors are given by 

\begin{eqnarray}
\left\{
\begin{array}{c} 
V_{1}(k)= v_{1,+} \otimes v_{2,+}
\\[0.4cm]
V_{2}(k)= v_{1,+} \otimes v_{2,-}
\\[0.4cm]
V_{1}(k)= v_{1,-} \otimes v_{2,+}
\\[0.4cm]
V_{3}(k)= v_{1,-} \otimes v_{2,-}
\end{array}  
\right.  ,                  
\end{eqnarray}
where $v_{1,\pm}$ and $v_{2,\pm}$ are eigenvectors of $U_{1}(k/2)$ and $U_{1}(k/2)$, respectively. Given eigenvalues and eigenvectors, one can write $\Psi_{T}(k)$ as 

\begin{eqnarray*}
\Psi_{T}(k) =U(k)^{T} \Psi_{0}(k) = \sum_{z=1}^{4} \Lambda_{z}^{T} \langle V_{z}(k), \Psi_{0}(k)\rangle V_{z}(k).
\end{eqnarray*}

By inverse Fourier transformation, the amplitude of the wave function of walker at the
position $x$ and the step $T$ will be found as

\begin{eqnarray}
\Psi_{T}(x)=\int_{0}^{2\pi} e^{-ikx} \sum_{z=1}^{4} \Lambda_{z}^{T} \langle V_{z}(k), \Psi_{0}(k)\rangle V_{z}(k)\frac{dk}{2\pi}, \label{PSI}
\end{eqnarray}
and the probability of finding walker at position $x$ at step $T$ is obtained by $p_{T}(x)=\|\Psi_{T}(x)\|^2$ where $\| \|$ denotes vector norm. 

For the case of identical sub-coins ($\theta=\gamma$) with $\theta \in (0,\pi/2)$, the eigenvalues are $\Lambda_{1}=\lambda_{1,+}^2$, $\Lambda_{2}=\Lambda_{3}=1$ and $\Lambda_{4}=\lambda_{1,-}^2$. These eigenvalues provide the possibility of using the method of stationary phase \cite{Liu2009,Liu2012} to have 

\begin{eqnarray*}
\Psi_{T}(x)\approx\int_{0}^{2\pi} e^{-ikx} (-1)^T \sum_{z=2}^{3} \Lambda_{z}^{T} \langle V_{z}(k), \Psi_{0}(k)\rangle V_{z}(k)\frac{dk}{2\pi}, \label{PSII}
\end{eqnarray*}

It is a matter of calculation to obtain the limiting probability of finding the walker at each position as \cite{Liu2012} 

\begin{eqnarray}
p(0) & = & \tan^2 \theta \mu_{1}^2 (|\beta_{1}|^2+|\beta_{4}|^2)+ \sec \theta \mu_{1} (|\beta_{2}|^2+|\beta_{3}|^2)+              \notag
\\
&  &
\tan \theta\mu_{1}^2 Re(\beta_{2}\bar{\beta}_{4}+\beta_{3}\bar{\beta}_{4}-\beta_{1}\bar{\beta}_{2}-\beta_{1}\bar{\beta}_{3})-              \notag
\\
&  &
2 \tan \theta \mu_{1} Re(\beta_{2}\bar{\beta}_{3}),              \notag
\end{eqnarray}

\begin{eqnarray}
p(x) \lvert_{x\geq1} & = &  \mu_{1}^{4 x} \tan^2 \theta [\mu_{2}^{2} |\beta_{1}|^2+|\beta_{2}|^2+|\beta_{3}|^2+\mu_{1}^2|\beta_{4}|^2-              \notag
\\
&  &
2 \mu_{2} Re(\beta_{1}\bar{\beta}_{2}+\beta_{1}\bar{\beta}_{3})+2 Re(\beta_{1}\bar{\beta}_{4}+\beta_{2}\bar{\beta}_{3})-              \notag
\\
&  &
2 \mu_{1} Re(\beta_{2}\bar{\beta}_{4}+\beta_{3}\bar{\beta}_{4})],              \notag
\end{eqnarray}

\begin{eqnarray}
p(x) \lvert_{x\leq -1} & = &  \mu_{1}^{-4 x} \tan^2 \theta [\mu_{1}^{2} |\beta_{1}|^2+|\beta_{2}|^2+|\beta_{3}|^2+\mu_{2}^2|\beta_{4}|^2+              \notag
\\
&  &
2 \mu_{1} Re(\beta_{1}\bar{\beta}_{2}+\beta_{1}\bar{\beta}_{3})+2 Re(\beta_{1}\bar{\beta}_{4}+\beta_{2}\bar{\beta}_{3})+              \notag
\\
&  &
2 \mu_{2} Re(\beta_{2}\bar{\beta}_{4}+\beta_{3}\bar{\beta}_{4})],           
\end{eqnarray}
in which $\mu_{1}= \sec \theta (1-\sin \theta)$ and $\mu_{2}=\sec \theta (1+\sin \theta)$.

If one considers $\beta_{2}=\beta_{3}=0$, the probability densities would reduce to 

\begin{eqnarray}
p(0) & = & \tan^2 \theta \mu_{1}^2 (|\beta_{1}|^2+|\beta_{4}|^2),              \notag	
\\[0.2cm]	
p(x) \lvert_{x\geq1} & = &  \mu_{1}^{4 x} \tan^2 \theta [\mu_{2}^{2} |\alpha_{1}|^2+\mu_{1}^2|\alpha_{4}|^2+2 Re(\alpha_{1}\bar{\alpha}_{4})],              \notag
\\[0.2cm]
p(x) \lvert_{x\leq -1} & = &  \mu_{1}^{-4 x} \tan^2 \theta [\mu_{1}^{2} |\alpha_{1}|^2+\mu_{2}^2|\alpha_{4}|^2+             \notag
\\
& & 2 Re(\alpha_{1}\bar{\alpha}_{4})].
\end{eqnarray}

Obtained probability densities confirm that only in case of $|\beta_{1}|^2=|\beta_{4}|^2$, probability densities of $p(x) \lvert_{x\geq1}$ in each position would be same as the ones in $p(x) \lvert_{x\leq -1}$ which results into symmetrical PDD. This is in agreement with our earlier results where walk with initial state of $\ketm{\psi_{1}}_{Int}$ \eqref{int1} and identical sub-coin was considered. In this case, $\beta_{1}=\cos \eta$ and $\beta_{4}=e^{I\phi}\sin \eta$ and symmetrical PDD was found for $\eta=\pi/4$. 

In contrast, for $\beta_{1}=\beta_{4}=0$, which corresponds to considering initial state of $\ketm{\psi_{2}}_{Int}$ \eqref{int2} with identical sub-coins, one can find

\begin{eqnarray}
p(0) & = & \sec \theta \mu_{1} (|\beta_{2}|^2+|\beta_{3}|^2)-2 \tan \theta \mu_{1} Re(\beta_{2}\bar{\beta}_{3}),              \notag
\\[0.2cm]
p(x) \lvert_{x\geq1} & = &  \mu_{1}^{4 x} \tan^2 \theta [|\beta_{2}|^2+|\beta_{3}|^2+2 Re(\beta_{2}\bar{\beta}_{3})],              \notag
\\[0.2cm]
p(x) \lvert_{x\leq -1} & = &  \mu_{1}^{-4 x} \tan^2 \theta [|\beta_{2}|^2+|\beta_{3}|^2+2 Re(\beta_{2}\bar{\beta}_{3})],
\end{eqnarray} 
which show that the symmetry of PDD is independent of $\beta_{2}$ and $\beta_{3}$. This is in agreement with our findings for the QW with initial state of $\ketm{\psi_{2}}_{Int}$ \eqref{int2} and identical sub-coin where symmetry was independent of changes in amplitudes of bases in initial state. It should be noted that for this case, $\beta_{2}=\cos \alpha$ and $\beta_{3}=e^{I\phi}\sin \alpha$.

In the limit of $\theta \rightarrow 0$ with $\beta_{2}=\beta_{3}=0$, the obtained probability density amplitudes will be   

\begin{eqnarray}
\lim\limits_{\theta  \rightarrow 0 }p(0) & = & O(\beta_{1}^2,\beta_{4}^2) \theta^2-2 O(\beta_{1}^2,\beta_{4}^2) \theta^3+              \notag
\\
&  &
\frac{8}{3} O(\beta_{1}^2,\beta_{4}^2) \theta^4 + ...,              \notag	
\\[0.2cm]	
\lim\limits_{\theta  \rightarrow 0 } p(x) \lvert_{x\geq1} & = &  O'(\beta_{1}^2,\beta_{4}^2) \theta^2 - O''(\beta_{1}^2,\beta_{4}^2,x) \theta^3+              \notag
\\
&  &
O'''(\beta_{1}^2,\beta_{4}^2,x^2) \theta^4 + ...,              \notag
\\[0.2cm]
\lim\limits_{\theta  \rightarrow 0 } p(x) \lvert_{x\leq -1} & = &  O'(\beta_{1}^2,\beta_{4}^2) \theta^2 + O''(\beta_{1}^2,\beta_{4}^2,x) \theta^3+              \notag
\\
&  &
O'''(\beta_{1}^2,\beta_{4}^2,x^2) \theta^4 + ... \text{ },
\end{eqnarray}
where $O$, $O'$, $O''$, $O'''$ different terms containing mentioned orders of $\beta_{i}$ and $x$.
Evidently, in this limit, the probability at the origin is almost zero whereas at the right and left hand sides' amplitudes, due to presence of $x$ as a factor, they would be non-zero and considerably larger. This indicates the presence of two considerable peaks at right and left hand sides' amplitudes. This is consistent with plotted diagrams in Fig. \ref{Fig1} and reported behavior in table \ref{T1} for initial state of \eqref{int1}. 

In contrast, for the limit of $\theta \rightarrow 0$ with $\beta_{1}=\beta_{4}=0$, probability density amplitudes would reduce to  

\begin{eqnarray}
\lim\limits_{\theta  \rightarrow 0 }p(0) & = & O(\beta_{2}^2,\beta_{3}^2) - O'(\beta_{2}^2,\beta_{3}^2) \theta+               \notag
\\
&  &
O'(\beta_{2}^2,\beta_{3}^2) \theta^2 + ...,	              \notag
\\[0.2cm]	
\lim\limits_{\theta  \rightarrow 0 } p(x) \lvert_{x\geq1} & = &  O'(\beta_{2}^2,\beta_{3}^2) \theta^2 - O''(\beta_{1}^2,\beta_{4}^2,x) \theta^3+              \notag
\\
&  &
O'''(\beta_{1}^2,\beta_{4}^2,x^2) \theta^4 + ...,              \notag
\\[0.2cm]
\lim\limits_{\theta  \rightarrow 0 } p(x) \lvert_{x\leq -1} & = &  O'(\beta_{2}^2,\beta_{3}^2) \theta^2 + O''(\beta_{1}^2,\beta_{4}^2,x) \theta^3+              \notag
\\
&  &
O'''(\beta_{1}^2,\beta_{4}^2,x^2) \theta^4 + ...,
\end{eqnarray}
which shows that probability density at origin becomes high while at right and left hand sides, the probability density amplitudes become almost zero except for very large (small) positions ($x$). This shows the presence of a Three-peaks-zone in PDD which is in agreement with plotted diagrams in 
Fig. \ref{Fig1} and reported behavior in table \ref{T1} for initial state of \eqref{int2}. 

Finally, for the case of non-zero $\beta_{z} \neq 0$ with $\theta \rightarrow 0$, the amplitudes of probability density are found as 

\begin{eqnarray}
\lim\limits_{\theta  \rightarrow 0 }p(0) & = & O(\beta_{2}^2,\beta_{3}^2) - O'(\beta_{1},\beta_{2},\beta_{3},\beta_{4}) \theta+                \notag
\\
&  &
O''(\beta_{1},\beta_{2},\beta_{3},\beta_{4}) \theta^2 + ...,               \notag       
\\[0.2cm]	
\lim\limits_{\theta  \rightarrow 0 } p(x) \lvert_{x\geq1} & = &  O'''(\beta_{1},\beta_{2},\beta_{3},\beta_{4}) \theta^2 -             \notag
\\
&  & 
O''''(\beta_{1},\beta_{2},\beta_{3},\beta_{4},x) \theta^3 + ...,             \notag            
\\[0.2cm]
\lim\limits_{\theta  \rightarrow 0 } p(x) \lvert_{x\leq -1} & = &  O'''(\beta_{1},\beta_{2},\beta_{3},\beta_{4}) \theta^2 +             \notag
\\
&  & 
O''''(\beta_{1},\beta_{2},\beta_{3},\beta_{4},x) \theta^3 + ...,
\end{eqnarray}
where $O'''(\beta_{1},\beta_{2},\beta_{3},\beta_{4},x)$ and $O''''(\beta_{1},\beta_{2},\beta_{3},\beta_{4},x)$ contain terms such as $(1-4x) Re(\beta_{1}(\beta_{2}+\beta_{3}))$ and $(1-4x) Re(\beta_{4}(\beta_{2}+\beta_{3}))$. Accordingly, there are non-zero probability densities at the origin and also in most right and left hand sides positions confirming a Three-peaks-zone behavior. In addition, the symmetry of PDD is only determined by $\beta_{1} \text{ and }\beta_{4}$ and it is independent of $\beta_{2} \text{ and }\beta_{3}$. These are in agreement with behaviors reported for walk with initial state of \eqref{int3} and identical sub-coins (see Fig. \ref{Fig3} and table \ref{T3}). The same could be done for the other limits such as $\theta \rightarrow \pi/2$ confirming our numerical simulation and presented tables for different behaviors.

\begin{acknowledgements}
SP would like to thank S. Wimberger, S. Barkhofen, and
L. Lorz for the helpful comments and discussions.
\end{acknowledgements}


\begin{thebibliography}{99}    
 
 \bibitem{Kempe}
 J. ~Kempe, Contemp.\ Phys.\ \textbf{44}, 307 (2003).
 
 
 \bibitem{Shenvi}
 N. ~Shenvi, J. ~Kempe, and K. ~B. ~Whaley, Phys.\ Rev.\ A\ \textbf{67}, 052307 (2003).
 

\bibitem{Ambainis3}
  A.~Ambainis, J.~Kempe and A.~Rivosh, Proc.\ 16th ACM-SIAM SODA, 1099, (2005).
  
 
\bibitem{Mohseni}
 M.~Mohseni, P.~Rebentrost, S.~Lloyd and A.~Aspuru-Guzik, J.\ Chem.\ Phys.\ \textbf{129}, 174106 (2008). 
 
 
 \bibitem{Lovett}
 N. ~B. ~Lovett, S. ~Cooper, M. ~Everitt, M. ~Trevers, and V. ~Kendon, Phys.\ Rev.\ A\ \textbf{81}, 042330 (2010).


\bibitem{Montero2017} 
  M. ~Montero, Phys.\ Rev.\ A\ \textbf{95} 062326 (2017).
  
 
 
\bibitem{Aspuru-Guzik} 
  A. ~Aspuru-Guzik, A. ~D. ~Dutoi, P. ~J. ~Love and M. ~Head-Gordon, Science\  \textbf{309}, 1704 (2005).    
  

\bibitem{Kitagawa}
 T. ~Kitagawa, M. ~S. ~Rudner, E. ~Berg and E. ~Demler, Phys.\ Rev.\ A \textbf{82}, 033429 (2010).
 
\bibitem{Panahiyan2019}
 S. Panahiyan and S. Fritzsche, Phys. Rev. A \textbf{100}, 062115 (2019). 
 
  
\bibitem{Dernbach}  
 S. ~Dernbach, A. ~Mohseni-Kabir, S. ~Pal, D. ~Towsley and M. ~Gepner, [arXiv:1801.05417]. 
  
 
\bibitem{Franco}     
 C. ~Di ~Franco and M. ~Paternostro, Phys.\ Rev.\ A\ \textbf{91}, 012328 (2015).
 
 
\bibitem{Innocenti}     
 L. ~Innocenti et al Phys.\ Rev.\ A\ \textbf{96}, 062326 (2017).  
  
	
\bibitem{Karski}
  M. ~Karski et al, Science\ \textbf{325}, 174 (2009).
  

\bibitem{Schreiber}
  A. ~Schreiber et al, Phys.\ Rev.\ Lett.\ \textbf{104}, 050502 (2010).
  

\bibitem{Zahringer}
  F. ~Zahringer, G. ~Kirchmair, R. ~Gerritsma, E. ~Solano, R. ~Blatt and C. ~F. ~Roos, Phys.\ Rev.\ Lett.\ \textbf{104}, 100503 (2010). 
  

\bibitem{Dadras}
  S. ~Dadras et al, Phys.\ Rev.\ Lett.\ \textbf{121}, 070402 (2018).
  

\bibitem{Barkhofen}  
  S. ~Barkhofen, L. ~Lorz, T. ~Nitsche, C. ~Silberhorn and H. ~Schomerus, [arXiv:1804.09496]. 
  
\bibitem{Lorz}  
  L. ~Lorz et al, [arXiv:1809.00591].  
  
  
\bibitem{Horodecki} 
  R. ~Horodecki et al, Rev.\ Mod.\ Phys.\ \textbf{81}, 865 (2009).
  

\bibitem{Muralidharan}   
  S. ~Muralidharan and P. ~K. ~Panigrahi, Phys.\ Rev.\ A\ \textbf{77}, 032321 (2008).  
   

\bibitem{Bouwmeester} 
  D. ~Bouwmeester et al, Nature\ \textbf{390}, 575 (1997).
  
\bibitem{Ekert}
  A. ~K. ~Ekert, Phys.\ Rev.\ Lett.\ \textbf{67}, 661 (1991).
  

\bibitem{Jozsa}
  R. ~Jozsa and N, ~Linden, [arXiv:quant-ph/0201143].


\bibitem{Hirsch} 
  F. ~Hirsch et al, Phys.\ Rev.\ Lett.\ \textbf{117}, 190402 (2016).
  
\bibitem{Fillettaz}  
M. ~Fillettaz, F. ~Hirsch, S. ~Designolle, and N. ~Brunner, Phys. Rev. A 98, 022115 (2018).     
  	

\bibitem{Venegas}
   S. ~E. ~Venegas-Andraca, J. ~L. ~Ball, K. ~Burnett and S. ~Bose, New\ J.\ Phys.\ \textbf{7}, 221 (2005).
   

\bibitem{Liu2009}
   C. ~Liu and N. ~Petulante, Phys.\ Rev.\ A\ \textbf{79}, 032312 (2009).
   
   
\bibitem{Liu2012}
   C. ~Liu, Quant.\ Info.\ Proc.\ \textbf{11}, 1193 (2012).
   
   
\bibitem{Singh}   
   S. Singh et al, J. Phys. Commun. \textbf{3}, 055008 (2019).   
 

\bibitem{Inui2004} 
   N. Inui and N. Konno, Physica A \textbf{253}, 133, (2004).

\bibitem{Inui2005}
   N. Inui, N. Konno and E. Segawa, Phys. Rev. E. \textbf{72}. 056112 (2005).  
  
 
\bibitem{Buarque}
  A. R. C. Buarque and W. S. Dias, Phys. Rev. A \textbf{101}, 023802 (2020).   
   
   
\bibitem{Brun}   
   T. ~A. ~Brun, H. ~A. ~Carteret and A. ~Ambainis. Phys.\ Rev.\ Lett.\ \textbf{91}, 130602 (2003).

\bibitem{kosik}
   J. ~Kosik, V. ~Buzek and M. ~Hillery, Phys.\ Rev.\ A\ \textbf{74}, 022310 (2006).

\bibitem{Kendon}
   V. ~Kendon, Math.\ Struct.\ in Comp.\ Sci.\ \textbf{17}, 1169 (2006).

\bibitem{Romanelli}
   A. ~Romanelli, Phys.\ Rev.\ A\ \textbf{76}, 054306 (2007).

\bibitem{Annabestani}
   M. ~Annabestani, S. ~J. ~Akhtarshenas and M. ~R. ~Abolhasani, Phys.\ Rev.\ A\ \textbf{81}, 032321 (2010).

\bibitem{Venegas-Andraca}
   S. ~E. ~Venegas-Andraca, Quant.\ Info.\ Process.\ \textbf{11}, 1015 (2012).

\bibitem{Alberti}   
   A. ~Alberti, W. ~Alt, R. ~Werner and D. ~Meschede, New\ J.\ Phys.\ \textbf{16}, 123052 (2014). 
   

\bibitem{Panahiyan} 
   S. ~Panahiyan and S. Fritzsche, New\ J.\ Phys.\ \textbf{20}, 083028 (2018). 
   
  
\bibitem{Childs2004}
   A. ~M. ~Childs and J. ~Goldstone, Phys.\ Rev.\ A\ \textbf{70}, 022314 (2004). 

\bibitem{Santha}
   M. ~Santha, [arXiv:0808.0059].  
   
 
\bibitem{Arnault}  
  P. ~Arnault, G. Di ~Molfetta, M. ~Brachet, and F. ~Debbasch, Phys. Rev. A \textbf{94}, 012335 (2016). 
    
  
\bibitem{Strauch}   
  F. W. ~Strauch, Phys. Rev. A \textbf{73}, 054302 (2006).    
   
  
\bibitem{Rajendran}
  J. ~Rajendran  and C. ~Benjamin, R.\ Soc.\ open\ sci.\ \textbf{5}, 171599 (2018). 
  
\bibitem{Machida}
  T. ~Machida, F. A. ~Grunbaum, Quant.\ Info.\ Process.\ \textbf{17}, 241 (2018).  
  
   
  
\bibitem{Kovlakov}  
  E. V. ~Kovlakov, S. S. ~Straupe, and  S. P. ~Kulik, Phys. Rev. A \textbf{98}, 060301 (2018). 
  
\bibitem{Giordani} 
  T. Giordani et al, Phys. Rev. Lett. \textbf{122}, 020503 (2019).   
  

\bibitem{Chandrashekar2008} 
  C. M. Chandrashekar et al, Phys. Rev. A \textbf{77}, 032326 (2008).   

\bibitem{Chandrashekar2011} 
  C. M. Chandrashekar and S. Banerjee, Phys. Lett. A \textbf{375}, 1553 (2011).     
  

  
\bibitem{Shannon}
  C. ~E. ~Shannon, The Bell System Technical Journal \textbf{27}, 623 (1948).


\bibitem{Nielsen}
  M. ~A. ~Nielsen and I. ~L. ~Chuang, \textit{Quantum computation and quantum information} (Cambridge University Press, 2010).  
   
   
\bibitem{Ide}
  Y. ~Ide, N. ~Konno and T. ~Machida, Quant.\ Inf.\ and Comput.\ \textbf{11}, 855 (2011).
   

\bibitem{Qiang}   
  X. ~Qiang et al, Nature\ Commun.\ \textbf{7}, 11511 (2016).    
    

\end{thebibliography}
\end{document}